\documentclass[aps,prx,reprint,superscriptaddress,footnote, twocolumn]{revtex4-1}

\usepackage{appendix}
\usepackage{caption}
\usepackage{subcaption}

\usepackage{tikz}  
\usetikzlibrary{arrows,shapes,positioning,shadows,backgrounds,fit}

\usepackage{relsize}
\usepackage{graphicx}
\usepackage{amsmath, bbm}
\usepackage{amssymb}
\usepackage{amsthm}
\usepackage{mathrsfs}
\usepackage{xcolor}
\usepackage{cancel}
\usepackage{titlesec}
\usepackage{enumitem}  
\usepackage{mathtools}
\usepackage{braket}
\usepackage[normalem]{ulem}

\titlespacing{\section}{0ex}{2ex}{0.4ex}
\titleformat{name=\section}{\bf}{\thesection.}{.3em}{}

\usepackage{soul}
\usepackage{dsfont}

\def\be{\begin{eqnarray}}
\def\ee{\end{eqnarray}}
\newcommand{\tr}[1]{\text{Tr}\left(#1\right)}

\theoremstyle{plain}

\definecolor{myblue}{rgb}{0.2,0.2,0.8}
\definecolor{myblack}{rgb}{0,0,0}
\definecolor{myurl}{rgb}{0.1,0.1,0.4}

\usepackage[colorlinks=true,citecolor=myblue,linkcolor=myblack,urlcolor=myurl]{hyperref}

\begin{document}

\title{Statistical Mechanics of Random Mixed State Ensembles with Fixed Energy}

\date{\today}

\author{Harry J.~D. Miller}
\affiliation{Department of Physics and Astronomy, The University of Manchester, Manchester M13 9PL, UK}

\begin{abstract}
Mixed state ensembles such as the Bures-Hall and Hilbert-Schmidt measure are probability distributions that characterise the statistical properties of random density matrices and can be used to determine the typical features of mixed quantum states. Here we extend this framework by considering the properties of random states with fixed average energy, and the ensemble-averaged density matrix is derived under this additional physical constraint. This gives rise to a type of microcanonical ensemble for random mixed states and we connect its properties to a statistical mechanical entropy and temperature. Our results are illustrated using a variety of simple spin systems, and we find that they can exhibit exotic features such as a first order phase transition at zero temperature in the absence of energetic interactions and finite relative energy fluctuations in the thermodynamic limit.
\end{abstract}

\maketitle

\section{Introduction}

\

Random quantum states  play an important role across many topics in quantum theory, including quantum information theory \cite{collins2016random,cross2019validating}, quantum chaos \cite{trail2008entanglement,ho2022exact}, black hole physics \cite{page1993information,kudler2021relative}, Bayesian state estimation \cite{schmied2016quantum}, and the foundations of statistical mechanics \cite{goldstein2006canonical,gogolin2016equilibration}. Whilst one may think of a random pure state as a uniformly sampled point on a projective Hilbert space \cite{wootters1990random}, the notion of a random state can be generalised to all mixed density matrices as well \cite{Hall1998,zyczkowski2003hilbert,Zyczkowski2005,Zyczkowski2011,Sommers2003,Sommers2004,nechita2007asymptotics,Sarkar2019,Sarkar2021}. The properties of such random density matrices are well studied and can be used to understand the typical properties of mixed quantum states such as their purity \cite{lubkin1978entropy,Zyczkowski2005}, subsystem entropy \cite{page1993average,bianchi2019typical,nadal2011statistical,wei2020exact}, distinguishability \cite{kudler2021relative}, and ergotropy \cite{hovhannisyan2024concentration}.

For many situations it is often unphysical to assume that a random quantum many-body system has access to its entire state space, and one can modify random ensembles under more realistic physical constraints \cite{garnerone2013generalized,hamma2012quantum,mark2024maximum}. One natural constraint is through energy conservation, which is a central feature connecting randomness to statistical mechanics. However, in quantum theory there are a number of approaches for introducing energy constraints to build a random ensemble. Standard ensembles are comprised of pure orthogonal eigenstates of a Hamiltonian $H$ and take the form of uniform probabilistic mixtures of states with definite energy \cite{landau2013statistical}. Alternatively if one assumes that the system is a random pure state $\ket{\psi}$ rather than an energy eigenstate, energy constraints can be introduced by restricting a uniform distribution of states to the manifold of states with fixed energy expectation value $E=\bra{\psi} H \ket{\psi}$ \cite{muller2011concentration}, giving rise to another type of microcanonical ensemble \cite{Brody2007,Brody2007a}. While conditions of ensemble equivalence can ensure predictions of these different ensembles converge in the thermodynamic limit, differences can still be observed for finite sized systems \cite{ji2011nonthermal}. In general, it remains an ongoing problem to understand which ensembles are most appropriate for describing equilibrated quantum systems under different physical constraints and settings.       

In this paper we will explore a new type of ensemble by relaxing one of the central postulates of statistical mechanics, namely the assumption of purity of the system's microstates \cite{reimann2007typicality,gogolin2016equilibration}. Instead we allow the microstates that contribute to the ensemble of the system to be \textit{any} valid quantum state $\rho$, subject to the average energy constraint $\tr{H\rho}=E$. In this context it is natural to utilise the random matrix theory of density matrices. There are a number of motivations for investigating the role of random density matrices in statistical mechanics. Physically, random mixed states can be simulated and studied in composite chaotic quantum systems \cite{znidaric2003fidelity,kubotani2008exact,tomsovic2018eigenstate,Sarkar2021}. Recently, random density matrix ensembles have also been shown to emerge naturally in isolated quantum systems that are initially mixed and then probed by unsharp measurements on a segment of their Hilbert space, a concept known as mixed state deep thermalisation \cite{yu2025mixed,sherry2025mixed}. In Bayesian approaches to statistical mechanics, ensembles are constructed based on ignorance of the system configuration \cite{jaynes1982rationale}, and density matrix ensembles can be utilised to assign an appropriate uniform prior in the absence of information \cite{buvzek1998reconstruction}. Some approaches to the foundations of quantum theory also argue that density matrices, as opposed to pure states, should be treated as the fundamental microstates of the theory \cite{page1986density,maroney2005density,chen2021quantum,chen2022uniform}. If one accepts this premise then from a purely statistical mechanical point of view there is justification to consider equilibrium ensembles constructed from distributions over the full set of density matrices. The final motivation concerns typicality in quantum theory, since energy-constrained mixed state ensembles provide a theoretical tool for determining the `average' properties of all quantum states confined to a given energy shell.     

Here we propose a mixed state ensemble for describing the average properties of all states under fixed energy, using tools from information geometry \cite{Hall1998,bengtsson2017geometry} to construct an underlying probability measure. This gives rise to a new type of microcanonical density matrix, and we show how to determine its properties from the assignment of a statistical mechanical entropy and temperature. Random matrix methods are used to derive general expressions for the density of states in terms of the Hamiltonian spectrum, and we investigate the properties of a some individual and many-body spin systems. Non-trivial properties of the microcanonical density matrix are shown to arise due to the underlying microstate space, including non-vanishing heat capacity at zero temperature, singular low-temperature behaviour in non-interacting systems, and violations of ensemble equivalence in the thermodynamic limit. 

\section{Geometric distributions of random mixed states}

\

In this section we first give an overview of how to define a statistical distribution of randomly sampled density matrices. To build such a measure the starting point is to consider the geometry of quantum state space \cite{bengtsson2017geometry}. Let $\mathcal{H}\simeq \mathbb{C}^d$ represent a finite dimensional Hilbert space and denote the accompanying set of quantum states, ie. the set of positive and unit-trace density matrices, by $\mathcal{S}(\mathcal{H})$. Each density matrix $\rho\in\mathcal{S}(\mathcal{H})$ can be viewed as a point within a $d^2-1$-dimensional real manifold. To obtain a random quantum state, one needs to sample these matrices uniformly with respect to a metric on $\mathcal{S}(\mathcal{H})$. Therefore we consider the pair $(\mathcal{M},g)$ as a $d^2-1$-dimensional, Riemannian manifold describing quantum states equipped with some metric $g$. The metric $g$ is derived from the infinitesimal distances between states, taking the generic form $ds^2=dL^2(\rho,\rho+d\rho)$ with $L(\rho_1,\rho_{2})$ a distance measure between states $\rho_1$ and $\rho_{2}$. Alongside the metric we denote $dV(\rho)$ as its Riemannian volume form which, upon normalisation, can be used as a uniform probability measure on $\mathcal{S}(\mathcal{H})$ \cite{Hall1998}. A subtle feature of the set of mixed states is that there is no unique choice of metric \cite{bengtsson2017geometry}; different functions may be chosen with different mathematical features and statistical properties. Before we specify any particular choice of the metric, consider the general parameterisation of a density matrix,
\begin{align}\label{eq:diag}
    \rho=UD U^\dagger, \ \ \ \ D=\text{diag}(r_0,r_{1},...r_{d-1})
\end{align}
where $U$ is a unitary matrix, $\sum_k r_k=1$ and $r_k\geq 0 \ \forall k$. We will only be concerned with metrics with a product volume form \cite{Zyczkowski2011},
\begin{align}\label{eq:haar}
dV(\rho)=d\mu_{\rho}(r_0,r_{1},..r_{d-1})\times d\mu_U,
\end{align}
which divides into a part dependent on the spectrum of $\rho$, multiplied by the Haar measure $d\mu_U$ on the set of unitary matrices $\mathcal{U}(d)$. While the latter is responsible for variations in the eigenvectors of the density matrices, the former governs the distribution of the eigenvalues. 

We will focus on two specific choices of measure commonly employed in studies of random quantum states. The simplest approach is to use the flat Hilbert-Schmidt distance $L_{HS}(\rho_1,\rho_{2})=(\tr{(\rho_1-\rho_{2})^2})^{1/2}$. This gives rise to a distribution $d\mu_\rho=d\mu_{HS}$ over the eigenvalues \cite{zyczkowski2003hilbert}
\begin{align}\label{eq:HS}
d\mu_{HS}(r_0,r_{1},..r_{d-1})\propto  dr_0 dr_{1} ...dr_{d-1}\prod_{\nu<\mu} (r_{\mu}-r_{\nu})^2,
\end{align}
which may be normalised for any finite dimension. We may interpret this as an induced measure \cite{zyczkowski2001induced}; in this case one takes a bipartite pure state uniformly sampled from an enlarged projective Hilbert space, then performs a partial trace to obtain a random reduced mixed state \cite{Zyczkowski2011}. Statistical properties of this distribution are well known, such as its average purity \cite{lubkin1978entropy} and entropy \cite{page1993average}. 

An alternative choice of measure can be defined from \textit{Bures length} \cite{HUBNER1992239,Petz1996b}, which takes the form
\begin{align}
L_B(\rho_1,\rho_{2}):=\big(2-2\sqrt{F(\rho_1,\rho_{2})}\big)^{1/2},
\end{align}
where
\begin{align}
F(\rho_1,\rho_{2})=\big\|\sqrt{\rho_1}\sqrt{\rho_{2}}\big\|_1^2
\end{align}
is the Uhlmann fidelity. Unlike the Hilbert-Schmidt distance, this measure may be interpreted as a quantum generalisation of the Fisher-Rao distance and is monotone under information processing \cite{petz1996monotone}. Moreover, it is the only monotone metric that also reduces to the unique Fubini-Study metric when restricted to the space of pure quantum states \cite{Petz1996b}. Considering the volume form for this metric, the resulting distribution over the eigenvalues is \cite{Hall1998}
\begin{align}\label{eq:hall}
d\mu_{BH}(r_0,r_{1},..r_{d-1})\propto  \frac{dr_0 dr_{1} ...dr_{d-1}}{\sqrt{r_0 r_{1}... r_{d-1}}}\prod_{\nu<\mu} \frac{(r_{\mu}-r_{\nu})^2}{r_{\nu}+r_{\mu}}
\end{align}
In the literature this measure is often referred to as the \textit{Bures-Hall ensemble} in random matrix theory, and it can also be normalised in finite dimensions \cite{Sommers2004}. It has been used to compute the typical values such as entropy, fidelity and level statistics \cite{Zyczkowski2005,forrester2016relating,Sarkar2019}, and there also exists a random matrix algorithm that can sample states according to~\eqref{eq:hall} \cite{AlOsipov2010}. Ultimately the Bures-Hall ensemble is less concentrated on highly mixed states, making it a more ideal candidate for sampling quantum states uniformly without any prior information \cite{Hall1998}.

\section{Microcanonical distribution for random density matrices}

\

While both the Hilbert-Schmidt~\eqref{eq:HS} and Bures-Hall distribution~\eqref{eq:hall} can be used to characterise the typical properties of randomly sampled states in $\mathcal{S}(\mathcal{H})$, they are not necessarily appropriate measures to describe many systems of interest since they do not account for physical constraints such as energy conservation. To remedy this we introduce a modified random matrix ensemble with an additional average energy constraint $\tr{H \rho}=E$ assigned to each state, where $H$ represents the Hamiltonian of the system. Energy eigenvalues of $H$ are labeled in increasing order, $E_0\leq  E_1\leq \cdots \leq E_{d-1}$ and may be degenerate. Assuming equal probability within the submanifold of states with fixed energy expectation, we introduce a normalised measure
\begin{align}\label{eq:mcmeasure}
    d\mu_E(\rho)=\omega^{-1}(E)\delta[E-\tr{H\rho}]dV(\rho)
\end{align}
with density of states 
\begin{align}\label{eq:doe}
    \omega(E):=\int_{\mathcal{S}(\mathcal{H})}dV(\rho) \ \delta[E-\tr{H\rho}].
\end{align}
This defines for us an energy-constrained version of either the Hilbert-Schmidt (HS) or Bures-Hall (BH) random matrix ensemble, depending on the choice of eigenvalue distribution. The linear statistics of the measure~\eqref{eq:mcmeasure} are quantified by the average state
\begin{align}\label{eq:mcstate}
    \bar{\rho}(E)=\int_{\mathcal{S}(\mathcal{H})}d\mu_E(\rho) \rho.
\end{align}
One may interpret this as a type of microcanonical equilibrium density matrix, and it has the natural property that it is stationary under Hamiltonian dynamics since 
\begin{align}
    [\bar{\rho}(E),H]=0,
\end{align}
which follows from the unitary invariance of $dV(\rho)$ (see Appendix~\ref{app:A}). However, it should be contrasted with other types of microcanonical state often used in statistical mechanics. For example, the standard definition of the microcanonical state consists of taking an equally weighted mixture of orthogonal energy eigenstates within a small energy window \cite{landau2013statistical}, with a density of states given by
\begin{align}
    \omega_{MC}(E)=\sum_k \delta(E-E_k).
\end{align}
Alternatively, the assumption of orthogonality can be relaxed to include all possible pure states within an energy shell \cite{Brody2007,Brody2007a,Campisi2013a,muller2011concentration,fine2009typical,Anza2020b,white2026eigenstate}. To do so one samples a random Haar pure state with measure $dV_{\text{Haar}}(\psi)$ and arrives at a density of states
\begin{align}\label{eq:pure}
    \omega_{\text{pure}}(E)=\int_{\mathcal{P}(\mathcal{H})}dV_{\text{Haar}}(\psi) \ \delta\bigg[E-\frac{\bra{\psi}H\ket{\psi}}{\braket{\psi}{\psi}}\bigg],
\end{align}
where $\mathcal{P}(\mathcal{H})$ denotes the projective Hilbert space. In any case these pure-state ensembles and their corresponding microcanonical density matrices are observably distinct from~\eqref{eq:mcstate}. Since $\bar{\rho}(E)$ is constructed from a random matrix ensemble, in this situation we should interpret each density matrix as a `\textit{microstate}' of the system rather than only the pure states.  

As a statistical mechanical system, the observable properties of the equilibrium state can be fully captured by its entropy function $S(E)$. To do so one can define a random-matrix analogue of the Gibbs volume entropy \cite{hilbert2014thermodynamic}, which is given by 
\begin{align}\label{eq:gibbs}
    S(E):=\text{ln} \ \Omega(E),
\end{align}
where 
\begin{align}
    \Omega(E):=\int^E_{E_0} dE' \ \omega(E'),
\end{align}
is the integrated density of states and $E_0$ the ground state energy of $H$. Alongside this we also have a notion of temperature associated with the state, which measures the rate of change of the microcanonical entropy with respect to energy,
\begin{align}\label{eq:temp}
    \frac{1}{T(E)}:=\frac{\partial S(E)}{\partial E}=\frac{\omega(E)}{\Omega(E)}\geq0.
\end{align}
This non-negativity follows from the use of the Gibbs volume entropy and will be assumed throughout. To see how the entropy and temperature characterise the properties of the density matrix~\eqref{eq:mcstate}, consider applying a small perturbation to the Hamiltonian, $H\mapsto H+\lambda \mathcal{O}$ with $\lambda$ a small parameter and $\mathcal{O}$ an observable of interest. Denote the new entropy of the system within an energy shell $E$ by $S_\lambda(E)$. It then follows (see Appendix~\ref{app:B}) that the expectation $\langle \mathcal{O}\rangle_E=\tr{\mathcal{O}\bar{\rho}(E)}$ of the original state is given by the simple perturbation formula
\begin{align}\label{eq:pk}
    \langle \mathcal{O}\rangle_E=-T(E)\frac{\partial  S_\lambda(E)}{\partial \lambda}\bigg|_{\lambda=0}. 
\end{align}
Therefore we can recognize that the volume entropy~\eqref{eq:gibbs} plays the role of the thermodynamic potential for the system and $T(E)$ fixes the energy scale. One can then determine any observable property of the equilibrium state from the static response of the entropy with respect to a perturbation. For example, to find the second moment in energy one can choose $\mathcal{O}=H^2$, as we will use later to compute the energy fluctuations of $\bar{\rho}(E)$.

\section{Two-level system}

\

\begin{figure}[!t]
  \centering
  \includegraphics[width=0.85\columnwidth]{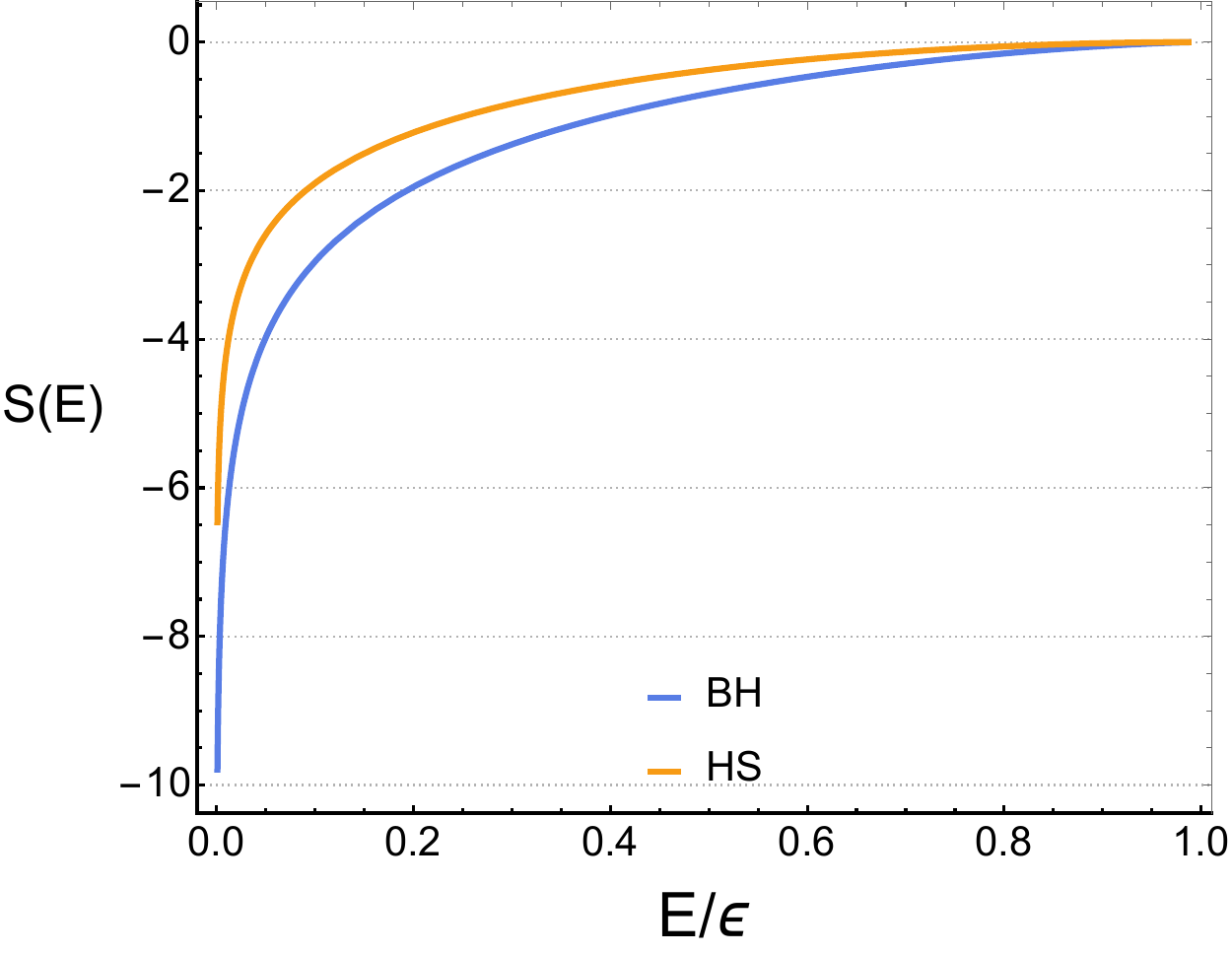}
  \caption{Volume entropy $S(E)$ for a two-level system with spacing $\epsilon$, plotted as a function of energy $E$. We compare the BH ensemble's entropy (blue) with the HS ensemble (orange).}
  \label{fig:BH_HS_qubit}
\end{figure}

As a warm up example we will first look at a two-level Hamiltonian 
\begin{align}
    H=\frac{\epsilon}{2}(\hat{\mathbb{I}}-\sigma_z),
\end{align}
whose eigenvalues are $0$ and $\epsilon$. This can be conveniently handled using the Bloch parameterisation 
\begin{align}
    \rho(r,\theta,\phi)=\frac{1}{2}(\hat{\mathbb{I}}+\vec{r}\cdot\vec{\sigma})
\end{align}
with polar coordinates $(r,\theta,\phi)$ and Bloch vector 
\begin{align}
    \vec{r}=\{r\sin\theta \cos\phi,\, r\sin\theta \sin\phi,\, r\cos\theta\}.
\end{align}
Each state carries an average energy expectation
\begin{align}
    \tr{H \rho}=\frac{\epsilon}{2}(1-r\cos\theta),
\end{align}
so the allowed energies satisfy $0\leq E\leq \epsilon$. The corresponding Bures-Hall volume element then takes a simple form \cite{Hall1998}
\begin{align}\label{eq:volume}
    dV_{BH}(r,\theta,\phi)=\frac{1}{8}\frac{r^2 \ \text{sin}\theta}{\sqrt{1-r^2}} \ dr d\theta d\phi, 
\end{align}
where $r\in[0,1], \ \theta\in[0,\pi], \ \phi\in[0,2\pi]$. To calculate the density of states one can employ the integral representation of the $\delta$-function, $\delta(x)=\frac{1}{2\pi}\int^\infty_{-\infty} d\lambda \ e^{-i\lambda x}$ and then find
\begin{align}\label{eq:omg2d}
    \nonumber\omega_{BH}(E)&\!=\!\int_{\mathbb{R}}  \frac{d\lambda \ e^{i\lambda E}}{16 \pi}  \int_{\mathbb{B}_3} dr d\theta d\phi \frac{r^2 \ \text{sin}\theta e^{-i\lambda\epsilon(1-r\cos\theta)/2}}{\sqrt{1-r^2}} , \\
    &\propto\int^\infty_{-\infty} d\lambda \ e^{i E \lambda}  \frac{J_1(\lambda\epsilon/2 )}{\lambda \epsilon}, 
\end{align}
where $J_n(x)$ is the Bessel function of the first kind. Completing the integration gives us an analytic expression for the integrated density of states,
\begin{align}\label{eq:qubitdos}
    \Omega_{BH}(E)\propto\bigg(\frac{2 E-\epsilon}{\epsilon^2}\bigg)\sqrt{\epsilon E-E^2}+\text{arcsin}\bigg(\sqrt{\frac{E}{\epsilon}}\bigg).
\end{align}
Let's now do the same calculation for the Hilbert-Schmidt volume form. This is just a Euclidean measure on the Bloch ball,
\begin{align}
    dV_{HS}(r,\theta,\phi)=r^2 \text{sin} \theta dr d\theta d\phi.
\end{align}
An analogous calculation of~\eqref{eq:omg2d} leads to a polynomial integrated density of states instead,
\begin{align}
    \Omega_{HS}(E)\propto 3\bigg(\frac{E}{\epsilon}\bigg)^2-2\bigg(\frac{E}{\epsilon}\bigg)^3,
\end{align}
which is equivalent, up to an overall multiplicative constant, to the cubic form obtained by direct integration on the interval $0\leq E\leq \epsilon$. In Figure~\ref{fig:BH_HS_qubit} the corresponding volume entropy is plotted as a function of energy for the two measures. While both are concave and monotonically increasing functions of energy, we see that there are slight discrepancies. 

Using the formula~\eqref{eq:pk} we can find the average density matrix for the BH ensemble, which is evidently diagonal in the energy basis $\{\ket{0},\ket{1}\}$ with matrix elements
\begin{align}\label{eq:qubit}
    \bar{\rho}(E)=\left(\begin{array}{cc}
          1-E/\epsilon & 0   \\ 
          0 & E/\epsilon
    \end{array}\right).
\end{align}
In fact, in this case the HS ensemble leads to \textit{exactly} the same average state. This coincidence is forced by the combination of stationarity, unit trace, and the single energy constraint $\tr{H\bar\rho(E)}=E$, which uniquely determine the diagonal entries in the two-level case. This equivalance breaks down once we consider higher dimensional systems, and is really just a quirk of the simple two-level state space. Differences between the average states will be observed when we consider further examples later on. 

\section{Calculating the density of states}

\

For larger dimensional systems, the central technical challenge is to evaluate the integral in the density of states~\eqref{eq:doe} over the state manifold $\mathcal{M}$. To do so it is first useful to introduce the canonical partition function, which is defined by the Laplace transform of the density of states,
\begin{align}\label{eq:partition}
    \nonumber\mathcal{Z}(\beta)&=\int^\infty_{E_0} dE \ \omega(E)e^{-\beta E}, \\
    &=\int_{\mathcal{S}(\mathcal{H})}d\mu_\rho\times d\mu_U \ e^{-\beta\tr{\hat{U}^\dagger\hat{H}\hat{U}\hat{D}}}. 
\end{align}
Once this is known we can derive the integrated density of states $\Omega(E)$ by inverting
\begin{align}\label{eq:lap_gen}
    \mathcal{Z}(\beta)=\beta\int^\infty_{0} dE \ e^{-\beta (E+E_0)} \Omega(E+E_0)
\end{align}
Using the volume form~\eqref{eq:haar} we perform the integral over the Haar measure first. For non-degenerate spectra this can be achieved using the \textit{Harish-Chandra-Itzykson-Zuber integral formula} \cite{mcswiggen2021harish}
\begin{align}\label{eq:canonical2}
    \mathcal{Z}(\beta)\propto \int_{0\leq r_0\leq...\leq r_{d-1}\leq 1} 
     d\mu_\rho \ \frac{\text{det} \ \big[e^{-\beta r_\nu  E_\mu}\big]_{\nu,\mu=0,...,d-1}}{\Delta(\rho)\Delta(-\beta H)}.
\end{align}
where $\Delta (\hat{A})=\prod_{\nu<\mu}(a_\mu-a_\nu)$ is the Vandermonde determinant, and the energy levels are arranged in increasing order. Degenerate spectra are then obtained by a standard coalescing-eigenvalue limiting procedure applied to the final formulas. Note that we are free to ignore any constant factors since these do not impact the observable properties of the microcanonical density matrix~\eqref{eq:mcstate}. This provides an expression only in terms of the energy spectrum. The remaining integral over the probabilities depends on the choice of measure, and we will now address the HS and BH ensembles separately. 

\subsection{Microcanonical Hilbert-Schmidt ensemble} 

\

For the HS measure, Andr\'eief’s integral formula \cite{forrester2019meet} can be used to perform the integral over the state eigenspectrum. The integral must be taken under the normalisation constraint $\sum_k r_k=1$, and to handle this we employ a Laplace transform technique outlined in \cite{Sarkar2019} to derive the following expression for the Hilbert-Schmidt partition function
\begin{align}\label{eq:HSpart1}
    \mathcal{Z}_{HS}(\beta)=\text{const}. \ \mathcal{L}_s^{-1}\bigg[\prod_k\frac{1}{( \beta E_k+s)^{n_k d}}\bigg](t)\bigg|_{t=1},
\end{align}
This formula is well defined even if there are degeneracies in the energy spectrum. The multiplicity of the level $E_k$ is denoted by the integer $n_k$, and the product over $k$ is now taken only over the $D\leq d$ \textit{distinct levels} $E_k\neq E_{k'}$. The presence of the simple poles in the denominator make the inverse Laplace transform amenable to the residue theorem, resulting in the following analytic expression:
\begin{align}\label{eq:HSpart2}
    \mathcal{Z}_{HS}(\beta)=\text{const}. \ \sum_{k} \sum_{m=0}^{n_k d-1} g_{k,m} \bigg[\frac{e^{-\beta E_k}}{\beta^{m+d(d-n_k)}}  \bigg],
\end{align}
with coefficients
\begin{align}
    \nonumber g_{k,m}&=\frac{(-1)^m}{(n_k d-m-1)!}  \\
    & \ \ \times\sum_{|\vec{m}|=m} \prod_{j\neq k}\binom{n_j d+m_j-1}{m_j}\frac{1}{(\Delta E_{jk})^{n_j d+m_j}}.
\end{align}
Here $\Delta E_{jk}=E_{j}-E_k$ is the energy gap and $|\vec{m}|=\sum_{j=1}^{D-1} m_j$ denotes a summation over the set of $D-1$ integers $m_j\in[0,m]$. A full derivation of~\eqref{eq:HSpart1} and subsequently~\eqref{eq:HSpart2} is provided in Appendix~\ref{app:C}.

To perform the remaining inverse Laplace transform in~\eqref{eq:lap_gen} we use the identity 
\begin{align}
    \mathcal{L}^{-1}_s[e^{-s A}/s^n](t):=\begin{cases}
    0 \ \ \ \ \ \ \ \ \ \ \ \ \ \ \ \ \ \        \text{for} \ t<A \\
    \frac{(t-A)^{n-1}}{(n-1)!}  \ \ \ \ \ \text{for} \ t\geq A \\
    \end{cases}
\end{align}
Therefore the integrated density of states is
\begin{align}
    \Omega_{HS}(E)=\text{const}.  \sum_{E_k\leq E} \sum_{m=0}^{n_k d-1}   \frac{g_{k,m} [E-E_k]^{m+d(d-n_k)}}{(m+d(d-n_k))!}
\end{align}
where we sum over all energy levels that satisfy $E\geq E_k$. From this we conclude that the integrated density $\Omega(E)$ is a piecewise polynomial function of energy $E$ of degree $d^2-1$. 

\subsection{Microcanonical Bures-Hall ensemble}

\

Finding the density of states for the BH measure requires slightly more involved methods, notably \textit{de Brujin’s integration theorem} \cite{de1955some} (see Appendix~\ref{app:D}). A lengthy derivation eventually leads to the following formula for the partition function:
\begin{align}
    \mathcal{Z}_{BH}(\beta)\propto \beta^{1-\frac{d^2}{2}}\mathcal{L}_s^{-1}\bigg[\prod^{d-1}_{\nu,\mu=0}\frac{1}{\sqrt{    E_\nu+s}+\sqrt{    E_\mu+s}}\bigg](\beta),
\end{align}
which is more difficult to handle analytically in comparison to the analogous HS version. To get the integrated density of states, it is more convenient to define the following generating function
\begin{align}\label{eq:GF}
G(s):=\text{const}.\prod^{d-1}_{\nu,\mu=0}\frac{1}{\sqrt{  \Delta E_\nu+s}+\sqrt{  \Delta E_\mu+s}}
\end{align}
where we denote the shifted levels $\Delta E_k=E_k-E_0$ and the product runs over the full spectrum including possible degeneracies. One can then show that the integrated density of states is related to $G(s)$ via a generalised Stieltjes transform of order $\alpha=\frac{d^2}{2}+1$:
\begin{align}\label{eq:result}
    G(s)=\int^\infty_{0} d E \ \frac{\Omega(E+E_0)}{(s+E)^{\alpha}}. 
\end{align}
For a given spectrum the Stieltjes transform can be inverted and expressed as a contour integral \cite{Schwarz2005generalized} 
\begin{align}\label{eq:CE}
    \Omega_{BH}(E)=\text{const}.\int_{\mathcal{C}_{\Delta E}} ds  (\Delta E+s)^{\frac{d^2}{2}-1}G(s),
\end{align}
where the contour $\mathcal{C}_{\Delta E}$ moves counterclockwise to and from the point $s=-\Delta E$ enclosing the origin with $\Delta E=E-E_0$. 

For a non-degenerate spectrum the contour integral can be handled in the following manner. Note that $G(s)$ has a branch cut along the negative real axis, and we may shrink the contour $\mathcal{C}_{\Delta E}$ so that it hugs either side of this cut up to the origin. This converts the contour into a real integral
\begin{align}\label{eq:stieltjes}
    \Omega(E)\propto\int^{\Delta E}_{0} ds \ (\Delta E-s)^{\frac{d^2}{2}-1}D(s),
\end{align}
where we now need to compute the discontinuity from crossing the branch,
\begin{align}
     D(s)&=\frac{1}{2\pi i}\lim_{\delta\to 0^+}\bigg(G(-s-i\delta)-G(-s+i\delta)\bigg). 
\end{align}
It is useful to divide the integral into segments connecting the points in the shifted spectrum $s=0, s=\Delta E_1, \cdots, \leq \Delta E$ that lie below the total energy shift. Within each interval one can determine how each term of the product making up $G(-s\pm i\delta )$ changes above and below the cut (see Appendix~\ref{app:E}), leading to the following formula
\begin{align}\label{eq:dosnumeric}
    \Omega(E)\propto\sum_{\Delta E_k\leq \Delta E}\int^{ \widetilde{\Delta E}_{k+1}}_{\Delta E_k} ds  (\Delta E-s)^{\frac{d^2}{2}-1}e^{-R_k(s)}\text{sin}  \theta_k(s),
\end{align}
where we define $\widetilde{\Delta E}_k=\text{min}\{ \Delta E, \Delta E_k\}$ and 
\begin{align}
    \nonumber R_k(s)&=\frac{1}{2}\sum_{j=0}^{d-1} \ln\!\big(\lvert \Delta E_j - s\rvert
      \big)+\sum_{\nu\leq k<\mu} \text{ln}(\Delta E_\mu-\Delta E_\nu) \\
    & \ \ \ \ +\sum_{\substack{\nu<\mu\\(\nu,\mu\le k)\lor(\nu,\mu>k)}}\!
2\ln\!\Bigl(\sqrt{\lvert \Delta E_\nu - s\rvert}
      +\sqrt{\lvert \Delta E_\mu - s\rvert}\Bigr) 
\end{align}
with phase
\begin{align}
    \theta_k(s)=\frac{(k+1)^2 \pi}{2}+2\sum_{\nu\leq k<\mu}\text{arctan}\bigg(\sqrt{\frac{|\Delta E_\nu-s|}{|\Delta E_\mu-s|}}\bigg).
\end{align}
In general these remaining integrals can be handled numerically. 

\section{Comparison of measures}

\

\begin{figure}[!t]
  \centering
  \includegraphics[width=\columnwidth]{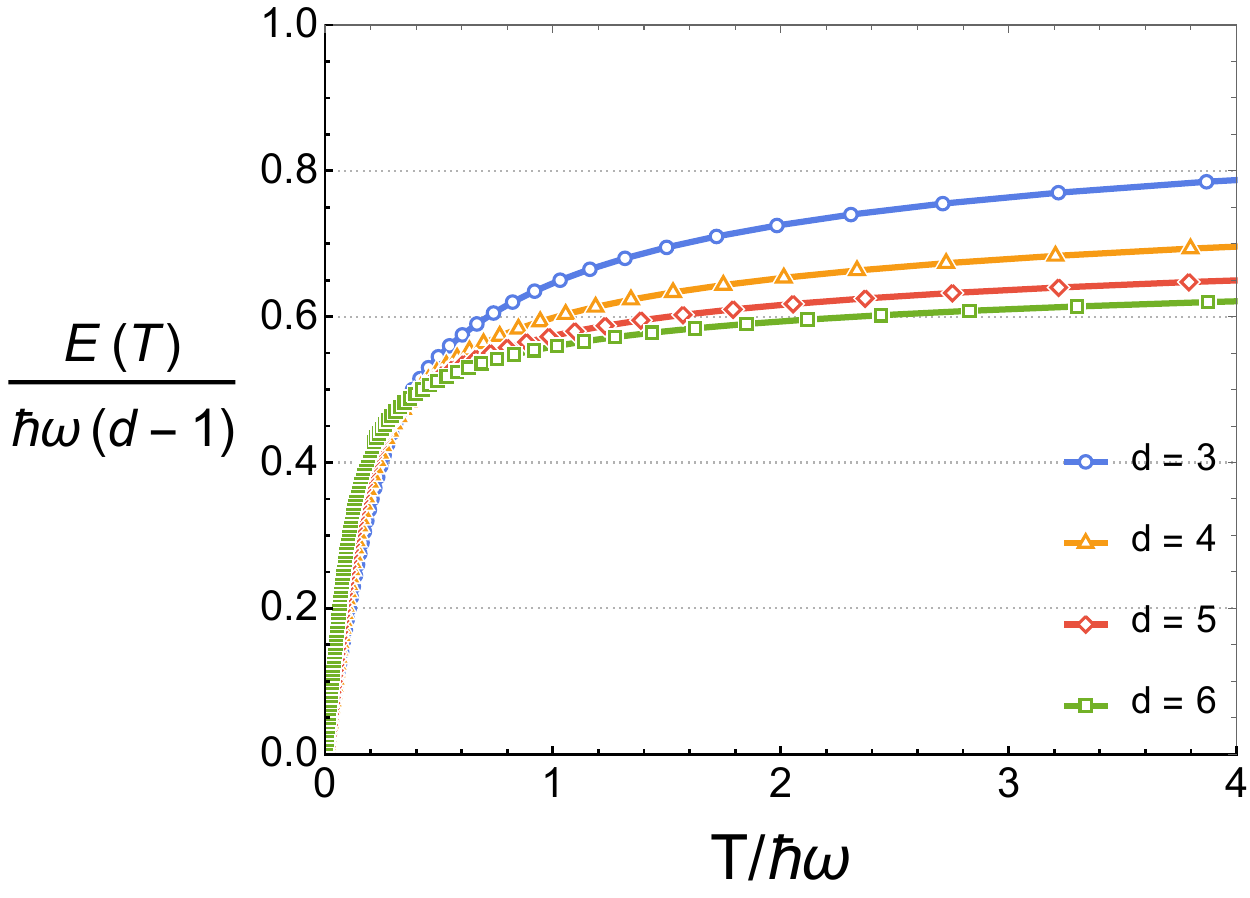}
  \caption{Average energy $E(T)$ as a function of temperature $T$ for the BH microcanonical ensemble with a linear spectrum.}
  \label{fig:bures_temp_lin}
\end{figure}

\begin{figure}[!t]
  \centering
  \includegraphics[width=\columnwidth]{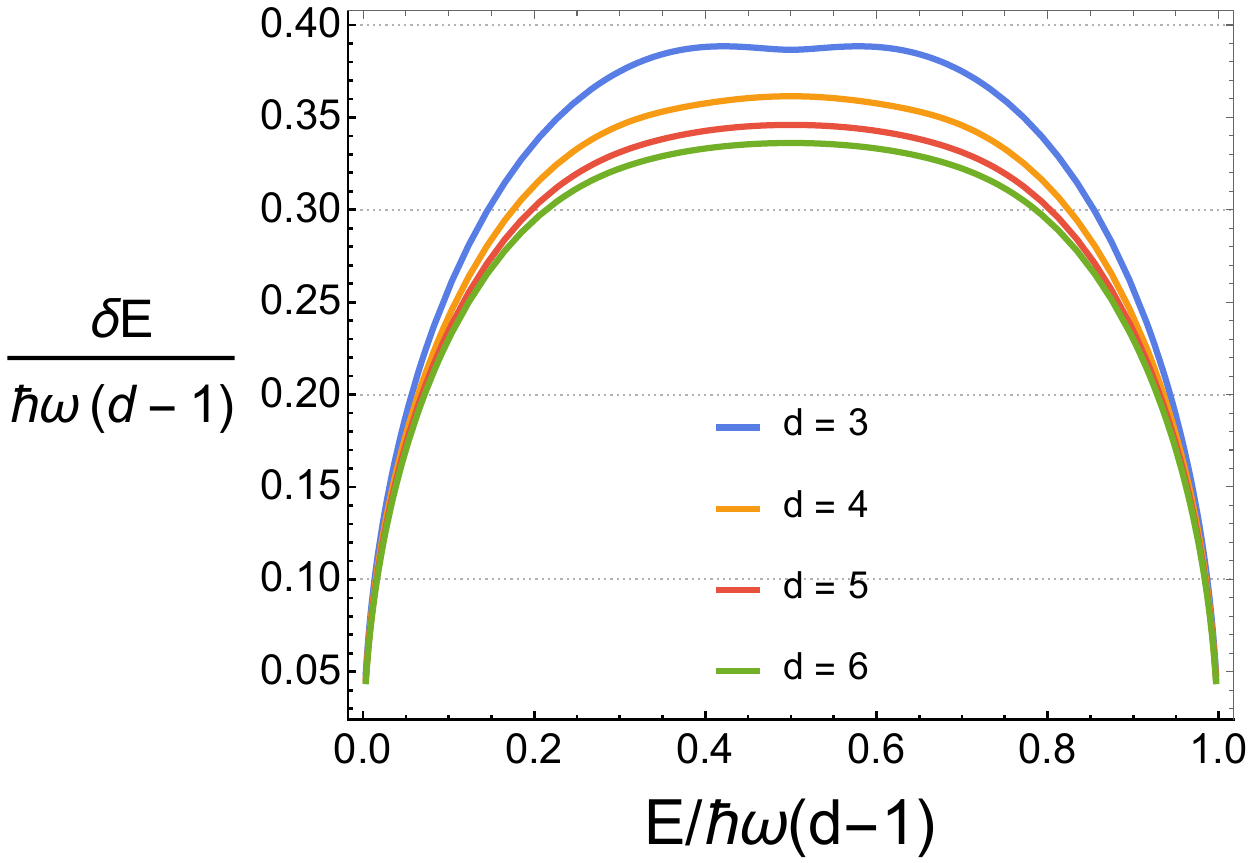}
  \caption{Energy fluctuations $\delta E$ as a function of average energy $E$ for the BH microcanonical ensemble with a linear spectrum.}
  \label{fig:bures_var_lin}
\end{figure}

So far we have derived general expressions for the density of states for the HS and BH microcanonical measures, and for simplicity we will first study the properties of a single spin system with a linear spectrum,
\begin{align}
    E_k=\hbar\omega k, \ \ k=0,1,\cdots , d-1.
\end{align}
Focusing first on the BH measure, in Figure~\ref{fig:bures_temp_lin} we plot the energy $E(T)$ as a function of the temperature for various system sizes. It can be seen that each system begins in the ground state at $T=0$, followed by a sharp monotonic increase in energy that eventually plateaus. In all cases the system will eventually reach its maximally excited state in the strict $T\to \infty$ limit. One can observe that the energy grows linearly with $T$ close to the ground state, indicating a finite heat capacity at absolute zero. In Figure~\ref{fig:bures_var_lin} we also plot the energy uncertainty $\delta E^2=\langle H^2 \rangle_E-E^2$ of the average state $\bar{\rho}(E)$ as a function of average energy $E$ for the same system sizes. As expected fluctuations go to zero as the system reaches either the ground or maximally excited energy eigenstate. While fluctuations tend to increase as the energy approaches the midpoint of the spectrum, monotonicity breaks down around the midpoint with minor oscillations in $\delta E$. The general trend is that the fluctuations decrease with increasing system size, though an analysis of the asymptotic behaviour is beyond the scope of this paper. 

\begin{figure}[!t]
  \raggedright
  \begin{subfigure}[b]{0.49\columnwidth}
    \includegraphics[width=\linewidth]{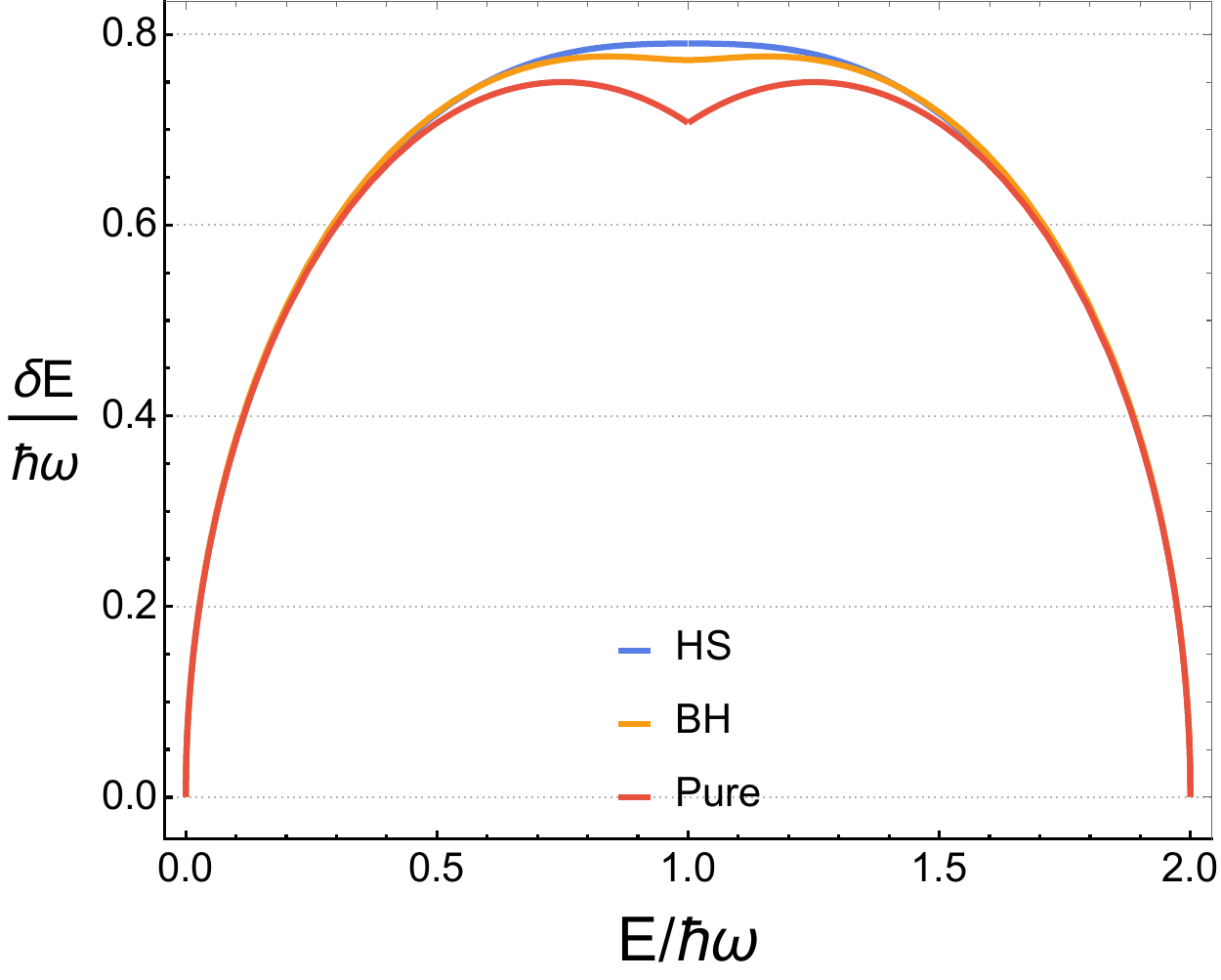}
    \subcaption{d=3}
    \label{fig:fluc_d3}
  \end{subfigure}%
  \hfill%
  \begin{subfigure}[b]{0.49\columnwidth}
    \includegraphics[width=\linewidth]{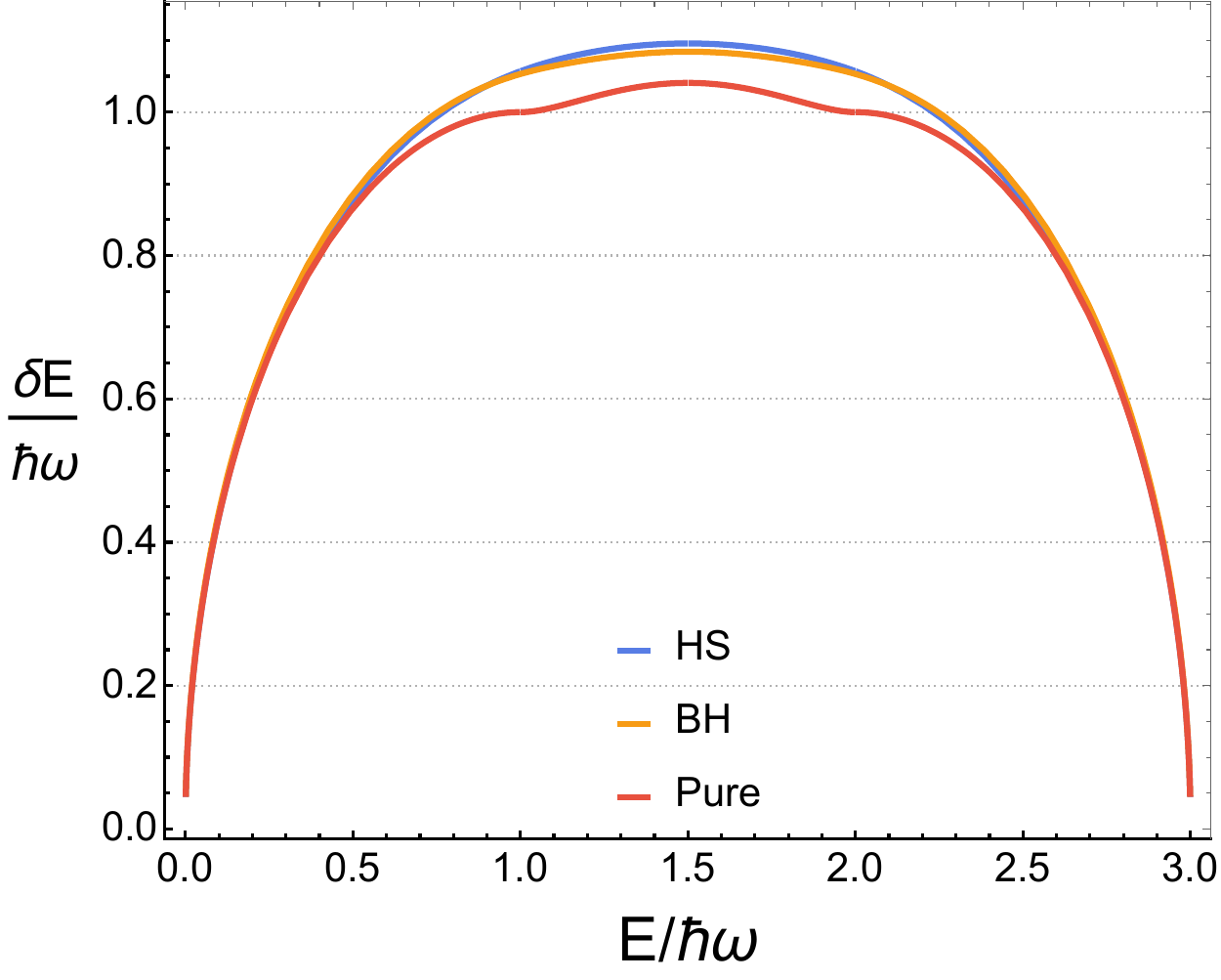}
    \subcaption{d=4}
    \label{fig:fluc_d4}
  \end{subfigure}

  \vspace{0.8em}

  \begin{subfigure}[b]{0.49\columnwidth}
    \includegraphics[width=\linewidth]{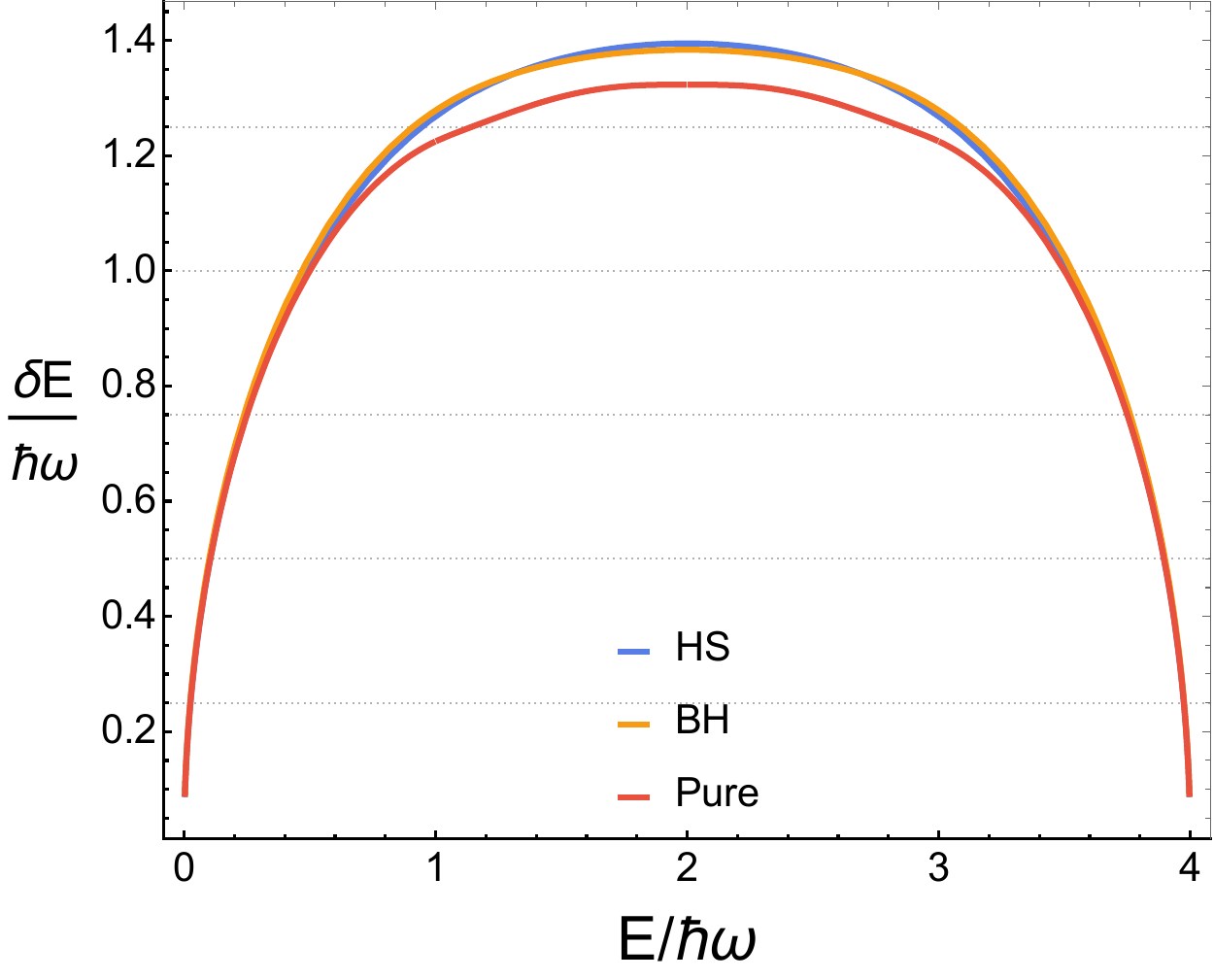}
    \subcaption{d=5}
    \label{fig:fluc_d5}
  \end{subfigure}%
  \hfill%
  \begin{subfigure}[b]{0.49\columnwidth}
    \includegraphics[width=\linewidth]{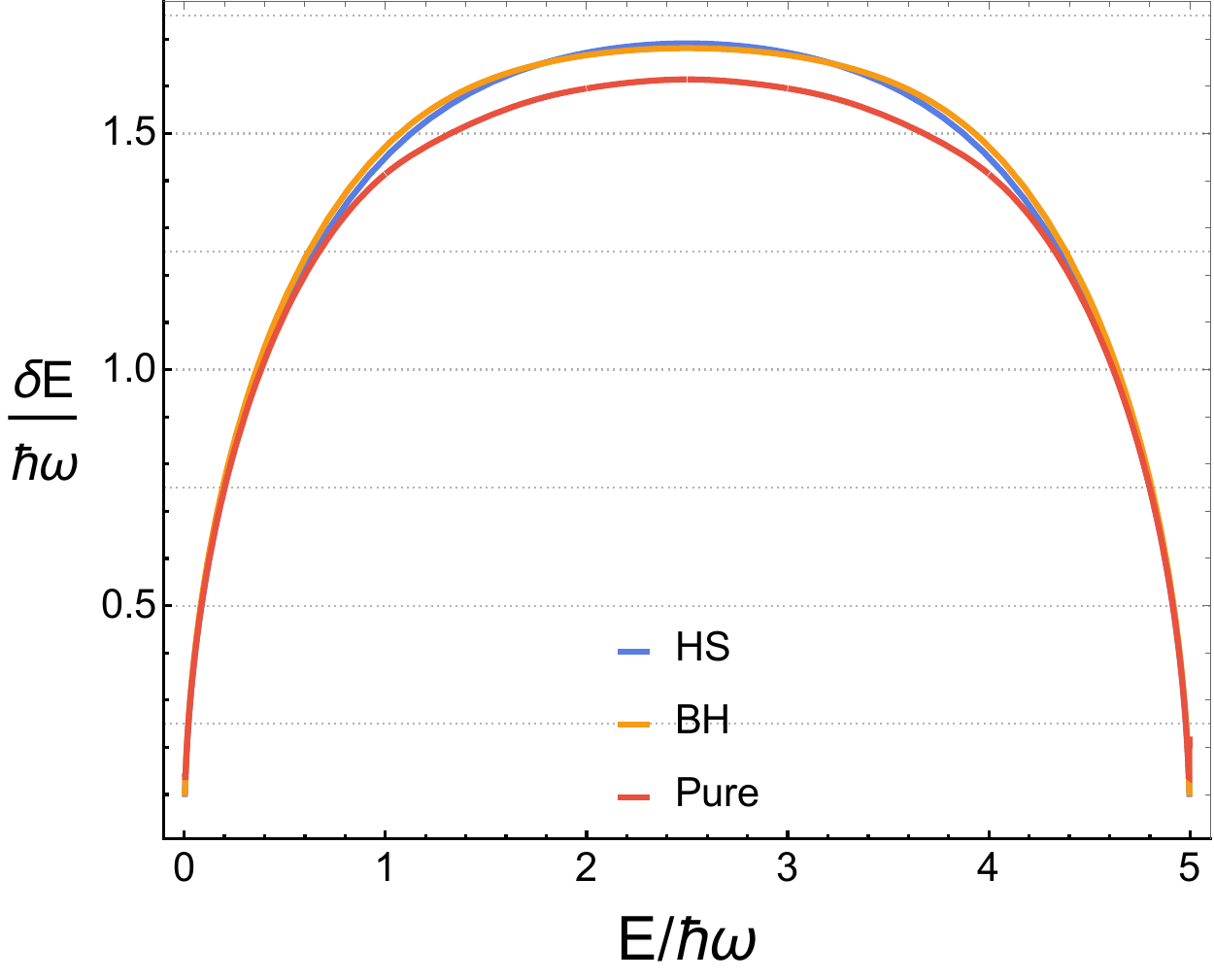}
    \subcaption{d=6}
    \label{fig:fluc_d6}
  \end{subfigure}

  \caption{Comparison of energy fluctuations $\delta E$ of the HS (blue), BH (orange) and pure-state (red) microcanonical density matrix with a linear spectrum at different dimensions $d$.}
  \label{fig:fluctuations_panel}
\end{figure}

In Figure~\ref{fig:fluctuations_panel} we compare the energy fluctuations of the HS microcanonical state with the BH state for various system sizes. We also introduce a further point of comparison with that of the pure-state microcanonical ensemble~\eqref{eq:pure}. One can see that the oscillations in $\delta E$ appear to be enhanced in the pure state ensemble, and lessened in the HS ensemble. Furthermore the magnitude of fluctuations for the pure ensemble is always lower than that of the HS and BH ensembles. The differences between all three tend to diminish with increasing system size, again suggesting ensemble equivalence in the thermodynamic limit for this example. In summary, this example confirms that for finite systems there are observable differences between these different ensembles that can be seen from the width of energy fluctuations.  

\section{Phase transition for non-interacting spins}

\

We will now investigate the statistical mechanics of a many-body system that has degenerate levels, which exhibits some unconventional statistical mechanical behaviour in the low energy regime. In particular we consider a chain of $N$ non-interacting spin-$1/2$ particles with total Hamiltonian $H=- \frac{B}{2}\sum^N_{i=1} \sigma_z^{(i)}$, where $B$ is an applied magnetic field and $\sigma_z^{(i)}$ the Pauli-$z$ matrix for the $i$'th spin. The Hilbert space dimension is $d=2^N$ and there are $N+1$ distinct energy levels $\{E_k \}_{k=0}^N$ each with multiplicity $n_k$, where 
\begin{align}
    E_k=-\bigg(\frac{N}{2}-k\bigg) B, \ \ \  n_k=\frac{N!}{k!(N-k)!}.
\end{align}
For the sake of brevity we will only consider the BH measure. In the low energy regime the microcanonical ensemble consists of a uniform distribution of density matrices that are energetically close to the ground state $E\sim E_0=-\frac{1}{2}N B$. We find, through an asymptotic expansion of the Laplace transform of $\Omega(E)$ (see Appendix~\ref{app:F} for details), that the ground state provides the dominant contribution to the contour integral~\eqref{eq:CE}. From this one arrives at the following approximation to the volume entropy:
\begin{align}\label{eq:Sexpand}
    \nonumber S(E)&\sim \frac{2^{2N}-1}{2}\text{ln} (\Delta E)+\frac{\Delta E}{(2^{2N+1}-2)}\bigg(\sum_{k=1}^N\frac{n_k}{\sqrt{\Delta E_k}}\bigg)^2 \\
    & \ \ \ \ \ \ \ \ \ \  -\sum_{\nu,\mu=1}^N n_\nu n_\mu\text{ln}(\sqrt{ \Delta E_\nu}+\sqrt{\Delta E_\mu})
\end{align}
where we have dropped terms of order $\mathcal{O}(\Delta E^2)$. From here it is straightforward to compute properties of the system. The two quantities we focus on are the mean energy per spin and relative energy fluctuations, defined by
\begin{align}
    \bar{\epsilon}_N(T):=\frac{E(T)}{N }, \ \ \ \Delta \bar{\epsilon}_N:=\frac{\sqrt{\langle H^2 \rangle_E-E^2}}{|E|}. 
\end{align}
To lowest order in temperature the mean energy per spin behaves as
\begin{align}\label{eq:PT}
     \bar{\epsilon}_N(T)
    \sim -\frac{B}{2}  +\frac{2^{2N}-1}{2N} T+\mathcal{O}(T^2).
\end{align}
As expected the system reaches the ground state at zero temperature. However, one also sees that the rate at which the mean energy approaches the ground state grows exponentially with system size. Consequently the cooling and subsequent loss of mixedness of the density matrix $\bar{\rho}(E)\mapsto \ket{E_0}\bra{E_0}$ exhibits  singular behaviour at zero-temperature  as one approaches the thermodynamic limit. This suggests a first order phase transition within a strict thermodynamic limit, which is unconventional behaviour for a non-interacting spin system and does not manifest in the standard canonical or microcanonical ensembles built from energy eigenstates. The origin of it stems from the non-extensive structure of the underlying random matrix ensemble, which includes contributions from all possible density matrices rather than just the separable energy eigenstates of the model. This ultimately means that, while the Hamiltonian itself is non-interacting, the ensemble-averaged state can still contain many-body correlations inherited from the allowed mixed-state microstates. We note that closely related singular low-temperature behaviour is also present for pure-state ensembles where the relevant correlations are generated by entanglement \cite{brody2001entanglement}. In the present context the larger mixed-state space encapsulates both entangled density matrices as well as classically correlated states, resulting in a greater rate of change in energy in~\eqref{eq:PT} with respect to the number of spins.

\begin{figure*}[!t]
  \centering
  \begin{subfigure}[b]{0.48\textwidth}
    \includegraphics[width=\linewidth]{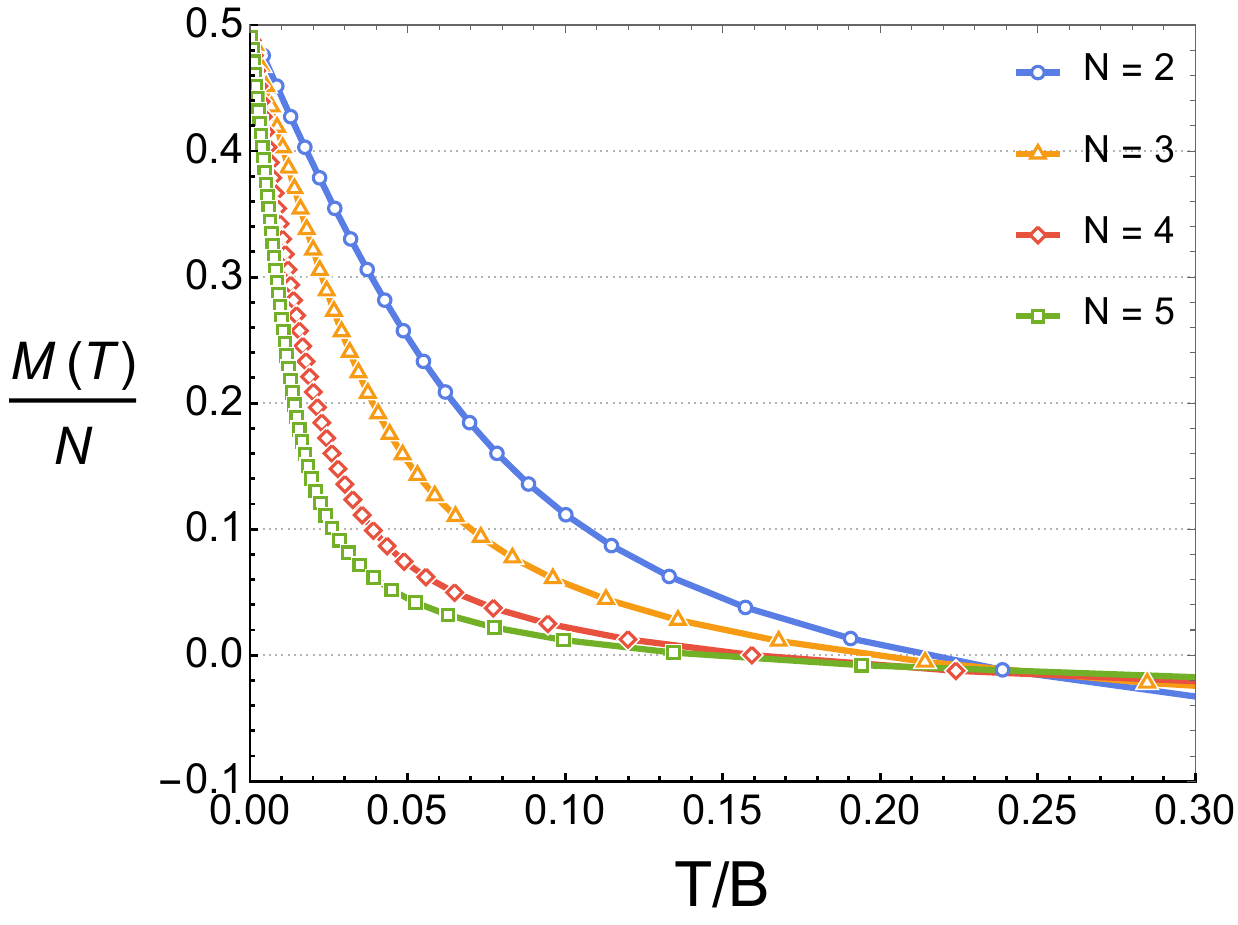}
    \caption{$J=0.2 B$}
    \label{fig:MagCW}
  \end{subfigure}
  \hfill
  \begin{subfigure}[b]{0.48\textwidth}
    \includegraphics[width=\linewidth]{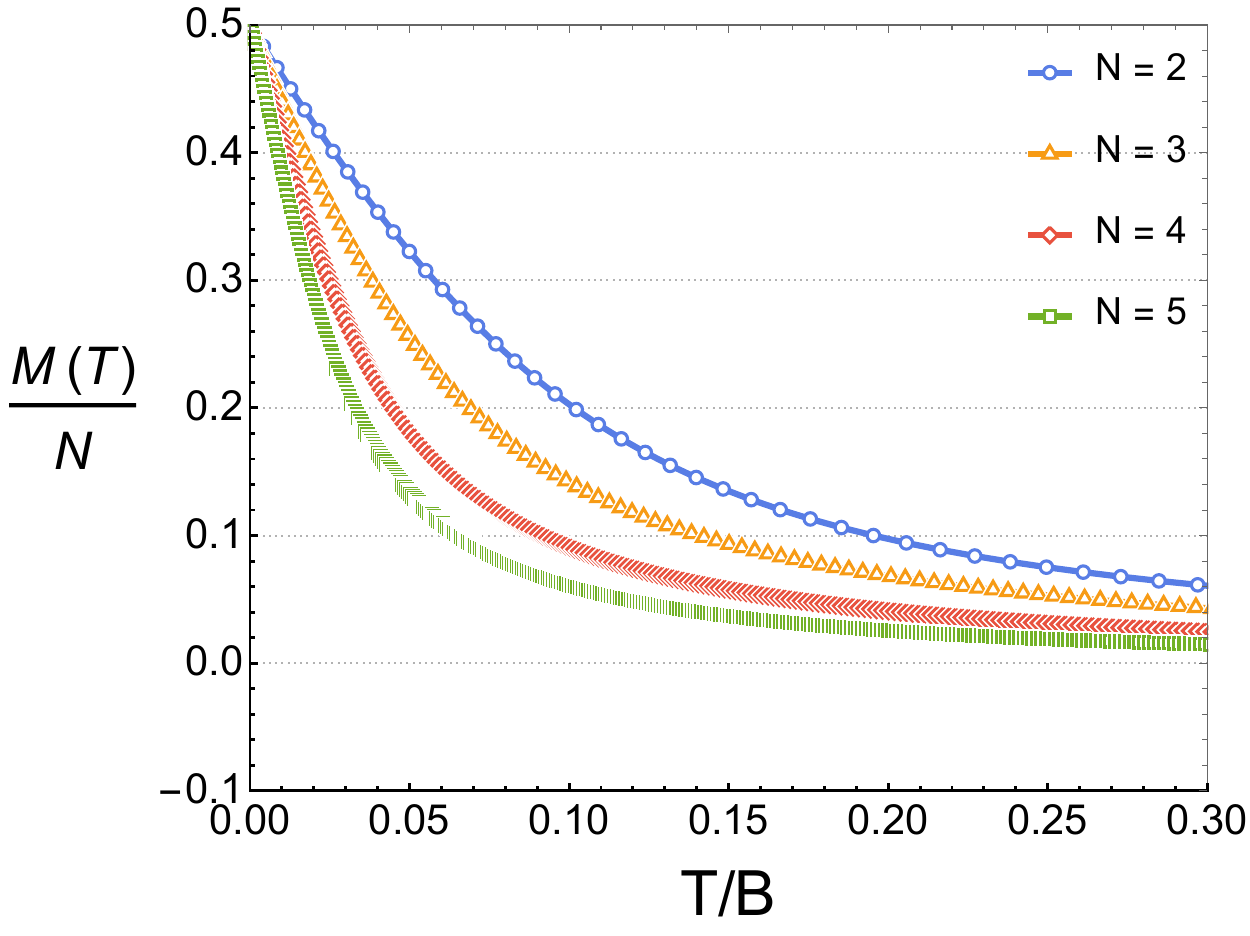}
    \caption{$J=20B$}
    \label{fig:MagCW2}
  \end{subfigure}
  \caption{Plots of the average magnetisation per spin of the Curie-Weiss model for the microcanonical HS density matrix as a function of temperature and for different system sizes. Plot (a) is weakly coupled $J=0.2B$ while plot (b) is strongly coupled, $J=20B$.}
\end{figure*}

The energy fluctuations also exhibit interesting behavior. Utilising the perturbation formula~\eqref{eq:pk} we find the large $N$ behaviour of the relative fluctuations to be (see Appendix~\ref{app:G})   
\begin{align}
    \Delta \bar{\epsilon}_N^2
    =-1-\frac{B}{2\bar{\epsilon}}
    +\frac{B(B+2\bar{\epsilon})}{16N\,\bar{\epsilon}^2}
    +\mathcal{O}(1/N^2)
\end{align}
The key feature is that this approaches a finite value in the thermodynamic limit,
\begin{align}
    \Delta \bar{\epsilon}_\infty=\sqrt{-1-\frac{B}{2\bar{\epsilon}}}, 
\end{align}
which is non-negative on the physical interval $-B/2\leq \bar{\epsilon}<0$ and vanishes only when the system reaches the ground state. The presence of \textit{macroscopic} energy fluctuations is another unusual property for a non-interacting spin system and indicates a violation of ensemble equivalence for this model. Again, the origin of this effect stems from the contributions of mixed states close to the ground state that are not energy eigenstates, and the influence of energy fluctuations continues to persist in the thermodynamic limit. 

\section{Interacting many-body system}

\

To round off our investigation we will now consider what happens when we include interactions between the spins. To make the analysis easier we will only focus on the HS ensemble as we do not anticipate significant qualitative differences to that of the BH measure. The example we look at is the Curie-Weiss Heisenberg model for $N$ spins \cite{brody2001entanglement}. The Hamiltonian is
\begin{align}
    H=-B S_z-\frac{J}{2N}\bigg(S^2-\frac{3N}{4}\bigg)
\end{align}
where $S_z=\frac{1}{2}\sum_{i=1}^N \sigma_z^{(i)}$ as before, $S$ is the total spin operator, $B>0$ is the magnetic field strength and $J$ denotes the interaction strength. The energy levels are
\begin{align}
    E_{s,m}=-Bm-\frac{J}{2N}\bigg(s(s+1)-\frac{3N}{4}\bigg)
\end{align}
where $s=s_{\min},s_{\min}+1,\cdots, \frac{N}{2}$ and $m=-s,\cdots,+s$. Here the minimum is set to the integer or half-integer
\begin{align}
    s_{\min}=\begin{cases}
        0 & \text{even } N, \\
        \frac{1}{2} & \text{odd } N.
    \end{cases}
\end{align}
The degeneracy of each level for a given total spin $s$ is
\begin{align}
    n_s=\binom{N}{\frac{N}{2}-s}-\binom{N}{\frac{N}{2}-s-1}.
\end{align}
We can determine the average magnetisation in the microcanonical density matrix from the derivative of the volume entropy with respect to the magnetic field strength, via the formula
\begin{align}
    M(E):=\tr{S_z \bar{\rho}(E)}=T(E)\frac{\partial S}{\partial B}.
\end{align}
In Figure~\ref{fig:MagCW} this is plotted as a function of the temperature, $M(T)$, for different system sizes for weak coupling $J=0.2 B$. We see that as the system is cooled down to its ground state the magnetisation rises to the expected value $M(T)=N/2$. The rate at which this occurs grows with increasing system size, similarly to what was observed in the non-interacting case. However, increasing the coupling strength as shown in Figure~\ref{fig:MagCW2} to $J=20 B$ indicates that this rate of change in $M(T)$ is reduced. Therefore increasing the interaction suppresses the singular ground-state crossover as one approaches the thermodynamic limit. 

\section{Discussion}

\

To summarise, we have used random density matrix distributions~\eqref{eq:haar} to construct a mixed-state ensemble for describing the uniform distribution of density matrices with a fixed energy expectation value, and developed mathematical methods for determining the average density matrix $\bar{\rho}(E)$ defined by~\eqref{eq:mcstate}. By interpreting this as a type of microcanonical equilibrium state, we introduced statistical mechanical notions of entropy and temperature and showed how these quantities directly relate to the observable properties of the system via the `\textit{thermodynamic relation}'~\eqref{eq:pk}. Unconventional statistical-mechanical phenomena linked to the mixed state-space were found in a simple non-interacting system, motivating further study of the properties of $\bar{\rho}(E)$ in more complex many-body systems.

First and foremost one may view $\bar{\rho}(E)$ as a theoretical tool to quantify average observable properties of random states subject to energy constraints. This generalises the familiar result that average expectation values of uniformly random states are simply given by the arithmetic mean of the observable's spectrum, $\langle \mathcal{O}\rangle=d^{-1}\sum^{d-1}_{k=0} \mathcal{O}_k$ \cite{venuti2013probability}. Our introduction of physical energy constraints is a key ingredient for understanding the typical properties of mixed quantum states in more realistic systems. While we have only focused on the linear properties of these random states, the next steps will be to explore the higher-order and non-linear properties of the random matrix ensemble~\eqref{eq:mcmeasure}, such as the purity and entanglement statistics \cite{AlOsipov2010,Sarkar2021}. The formalism here could be useful in quantum thermodynamics, eg. by providing a mixed-state generalisation of pure-state ensemble approaches to non-equilibrium fluctuation theorems \cite{Campisi2013a,Anza2020b}, as well as a tool for quantifying typical ergotropy of energy-constrained random states \cite{hovhannisyan2024concentration} and work extraction from unknown quantum states \cite{vsafranek2023work}. There are a number of ways to generalise our construction, the first of which would be to incorporate additional conserved quantities into the ensemble. Secondly, while we have only considered the microcanonical ensemble in this paper, one could also explore the properties of the canonical ensemble with a Gibbs weighting $d\mu_G(\rho)\propto e^{-\beta \tr{H \rho}}dV(\rho)$. Lastly, alternative mixed-state ensembles could be considered such as a recently proposed entropy-based ensemble \cite{miller2025entropy}.     

We conclude with a number of key open questions for future investigation:
\begin{itemize}
    \item \textit{Is there an algorithm that can prepare energy-constrained random states?} 
\end{itemize}
As previously mentioned, the procedure for generating unconstrained random density matrices distributed by either the Hilbert-Schmidt or Bures-Hall ensemble is well established, and can be achieved through combined global and local random unitary operations applied to a purified state in an enlarged Hilbert space \cite{Zyczkowski2011}. Naturally we would like to understand how to generate random states within a given energy shell manifold. Algorithms for sampling pure quantum states from fixed energy manifolds are known \cite{muller2011concentration} and could potentially be adapted to sample from the Bures-Hall ensemble. 
\begin{itemize}
    \item \textit{Are there classes of Hamiltonian that recover ensemble equivalence with the standard microcanonical ensemble in the thermodynamic limit?}
\end{itemize}
The simple systems that we have looked at in this paper show clear deviations from standard statistical mechanical predictions. However, it is conceivable that some classes of integrable and chaotic Hamiltonians could still reproduce the usual predictions of the quantum microcanonical density matrix in the thermodynamic limit. 
\begin{itemize}
    \item \textit{Can energy-constrained mixed-state ensembles describe equilibration in some many-body systems?}
\end{itemize}
As a counterpoint to the question of ensemble equivalence, it would be interesting to investigate if there are systems where the state $\bar{\rho}(E)$ provides more accurate predictions for the long-time behaviour of certain chaotic many-body systems. Recent studies of so-called deep thermalisation \cite{choi2023preparing,ippoliti2022solvable,lucas2023generalized,bhore2023deep,mark2024maximum} and Hilbert space ergodicity \cite{pilatowsky2023complete,logaric2025hilbert} suggest that energy-constrained pure state ensembles can approximate the behaviour of chaotic many-body systems that are projectively measured on part of their Hilbert space. More recently, assumptions about initial purity and sharpness of measurements have been relaxed \cite{yu2025mixed,sherry2025mixed}, resulting in the formation of mixed-state ensembles related to the random Hilbert-Schmidt distribution~\eqref{eq:HS}. Mixed-state ensembles such as the one considered here could play a role when constraints such as average energy conservation are imposed in these setups. A second possibility for the emergence of energy-constrained mixed states could be through Ehrenfest dynamics \cite{alonso2011statistics}, which is a known mechanism for equilibration to energy-constrained pure ensembles. 

\textit{Acknowledgments:} I thank Jake Iles-Smith, Karen Hovhannisyan and Christopher D. White for helpful discussions. I acknowledge funding from a Royal Society Research Fellowship (URF/R1/231394).

\bibliographystyle{apsrev4-1}
\bibliography{mybib2.bib}

\appendix

\widetext

\section{Stationarity of the ensemble}\label{app:A}

\

Consider the unitary process up to time $t\in\mathbb{R}$,
\begin{align}
    U_t=e^{i t H}.
\end{align}
The ensemble evolves
\begin{align}
    \nonumber U_t \bar{\rho}(E) U^\dagger_t&=\omega^{-1}(E)\int_{\mathcal{S}(\mathcal{H})}dV(\rho) \  \delta[E-\tr{H\rho}]U_t \rho U^\dagger_t, \\
    \nonumber&=\omega^{-1}(E)\int_{\mathcal{S}(\mathcal{H})}dV(\rho) \  \delta[E-\tr{U_t H U_t^\dagger U_t\rho U_t^\dagger}]U_t \rho U^\dagger_t, \\
    \nonumber&=\omega^{-1}(E)\int_{\mathcal{S}(\mathcal{H})}dV(U_t \rho U_t^\dagger) \  \delta[E-\tr{ H  U_t\rho U_t^\dagger}]U_t \rho U^\dagger_t, \\
    &=\bar{\rho}(E),
\end{align}
where we used the fact that $[U_t,H]=0$ in the third line, along with the unitary invariance of the measure, $dV(\rho)=dV(U \rho U^\dagger)$. Since the above holds for any $t\in\mathbb{R}$ we conclude $\big[\bar{\rho}(E),H\big]=0$. 

\section{Microcanonical expectation values}\label{app:B}

\

By linearity it follows that the expectation value $\langle \mathcal{O} \rangle_E=\tr{\bar{\rho}(E)\mathcal{O}}$ of some observable $\mathcal{O}$ with respect to the average density matrix is given by
\begin{align}
    \langle \mathcal{O} \rangle_E
    = \int_{\mathcal{S}(\mathcal{H})}d\mu_E(\rho)\,\tr{\mathcal{O}\rho}
    =\omega^{-1}(E)\int_{\mathcal{S}(\mathcal{H})}dV(\rho)\,
    \delta\!\big[E-\tr{H\rho}\big]\tr{\mathcal{O}\rho}.
\end{align}
We will derive a relationship between this average and the volume entropy of the system by considering the perturbation $H\mapsto H+\lambda \mathcal{O}$ with $\lambda$ a small scalar variable. The corresponding density of states is
\begin{align}
    \omega_\lambda(E)=\int_{\mathcal{S}(\mathcal{H})}dV(\rho)\,
    \delta\!\big[E-\tr{(H+\lambda\mathcal{O})\rho}\big],
\end{align}
and the associated integrated density of states and volume entropy are
\begin{align}
    \Omega_\lambda(E)=\int_{E_0(\lambda)}^E dE' \ \omega_\lambda(E'),
    \qquad
    S_\lambda(E)=\ln \Omega_\lambda(E),
\end{align}
where $E_0(\lambda)$ is the ground-state energy of the perturbed Hamiltonian.

To proceed, differentiate $\Omega_\lambda(E)$ with respect to $\lambda$. By Leibniz' rule,
\begin{align}
    \partial_\lambda \Omega_\lambda(E)
    =-E_0'(\lambda)\,\omega_\lambda\!\big(E_0(\lambda)\big)
    +\int_{E_0(\lambda)}^E dE'\,\partial_\lambda \omega_\lambda(E').
\end{align}
For the ensembles considered here the density of states vanishes at the lower spectral edge,
\begin{align}
    \omega_\lambda\!\big(E_0(\lambda)\big)=0,
\end{align}
since the set of states satisfying $\tr{(H+\lambda\mathcal{O})\rho}=E_0(\lambda)$ consists only of ground-state density matrices and has zero Riemannian volume inside the full mixed-state manifold. Hence the boundary term vanishes and
\begin{align}
    \partial_\lambda \Omega_\lambda(E)
    =\int_{E_0(\lambda)}^E dE'\,\partial_\lambda \omega_\lambda(E').
\end{align}
Using the identity
\begin{align}
    \partial_\lambda \delta\!\big[E'-\tr{(H+\lambda\mathcal{O})\rho}\big]
    =
    -\tr{\mathcal{O}\rho}\,
    \delta'\!\big[E'-\tr{(H+\lambda\mathcal{O})\rho}\big],
\end{align}
we obtain
\begin{align}
    \partial_\lambda \omega_\lambda(E')
    =
    -\int_{\mathcal{S}(\mathcal{H})}dV(\rho)\,
    \tr{\mathcal{O}\rho}\,
    \delta'\!\big[E'-\tr{(H+\lambda\mathcal{O})\rho}\big].
\end{align}
Substituting this into the derivative of $\Omega_\lambda(E)$ and integrating over $E'$ gives
\begin{align}
    \nonumber
    \partial_\lambda \Omega_\lambda(E)
    &=
    -\int_{\mathcal{S}(\mathcal{H})}dV(\rho)\,\tr{\mathcal{O}\rho}
    \int_{E_0(\lambda)}^E dE'\,
    \delta'\!\big[E'-\tr{(H+\lambda\mathcal{O})\rho}\big] \\
    \nonumber
    &=
    -\int_{\mathcal{S}(\mathcal{H})}dV(\rho)\,\tr{\mathcal{O}\rho}
    \Big[
    \delta\!\big(E-\tr{(H+\lambda\mathcal{O})\rho}\big)
    -
    \delta\!\big(E_0(\lambda)-\tr{(H+\lambda\mathcal{O})\rho}\big)
    \Big].
\end{align}
The second term vanishes for the same reason as above: it is supported only on ground-state density matrices, which form a zero-volume subset of $\mathcal{S}(\mathcal{H})$. Therefore
\begin{align}
    \partial_\lambda \Omega_\lambda(E)
    =
    -\int_{\mathcal{S}(\mathcal{H})}dV(\rho)\,\tr{\mathcal{O}\rho}\,
    \delta\!\big[E-\tr{(H+\lambda\mathcal{O})\rho}\big].
\end{align}
Evaluating at $\lambda=0$ then yields
\begin{align}
    \partial_\lambda \Omega_\lambda(E)\big|_{\lambda=0}
    =
    -\int_{\mathcal{S}(\mathcal{H})}dV(\rho)\,\tr{\mathcal{O}\rho}\,
    \delta\!\big[E-\tr{H\rho}\big].
\end{align}
Comparing this with the definition of $\langle \mathcal{O}\rangle_E$, we arrive at
\begin{align}
    \langle \mathcal{O}\rangle_E
    =
    -\frac{\partial_\lambda \Omega_\lambda(E)|_{\lambda=0}}{\omega(E)}.
\end{align}
Finally, using $\Omega_\lambda(E)=e^{S_\lambda(E)}$ together with
\begin{align}
    \omega(E)=\frac{\partial \Omega(E)}{\partial E}
    =e^{S(E)}\frac{\partial S(E)}{\partial E},
\end{align}
we find
\begin{align}
    \nonumber
    \langle \mathcal{O}\rangle_E
    &=
    -\frac{e^{S(E)}\,\partial_\lambda S_\lambda(E)|_{\lambda=0}}
    {e^{S(E)}\,\partial_E S(E)} \\
    &=
    -T(E)\,\partial_\lambda S_\lambda(E)\big|_{\lambda=0},
\end{align}
where we used the definition of temperature
\begin{align}
    \frac{1}{T(E)}=\frac{\partial S(E)}{\partial E}.
\end{align}
Therefore the expectation value of any observable may be obtained from the response of the volume entropy under the perturbation $H\mapsto H+\lambda\mathcal{O}$:
\begin{align}
    \langle \mathcal{O}\rangle_E
    =
    -T(E)\frac{\partial S_\lambda(E)}{\partial\lambda}\bigg|_{\lambda=0}.
\end{align}

\section{Deriving the density of states: Hilbert-Schmidt measure}\label{app:C}

\

For the HS measure, substitute~\eqref{eq:HS} into~\eqref{eq:canonical2}:
\begin{align}\label{eq:canonical3}
    \mathcal{Z}(\beta)\propto \frac{1}{\Delta(-\beta H)}\int_{0\leq r_0\leq...\leq r_{d-1}\leq 1} 
     dr_0\cdots dr_{d-1} \ \delta[\tr{D}-1] \ \Delta(\rho)\ \text{det} \ \big[e^{-\beta r_\nu  E_\mu}\big]_{\nu,\mu=0,...,d-1} .
\end{align}
It is instructive to recall that the Vandermonde determinant can be expressed as \cite{mehta2004random}
\begin{align}\label{eq:vand}
    \Delta(A)=\text{det}[a_\nu^{\mu} ]_{0\leq \nu,\mu\leq d-1}=\prod_{0\leq \nu<\mu\leq d-1}(a_\mu-a_{\nu}).
\end{align}
In order to complete the integral in~\eqref{eq:canonical3}, we replace the normalisation condition $\delta[\tr{D}-1]$ with a dummy variable $\delta[\tr{D}-t]$ and take a Laplace transform with respect to $t$:
\begin{align}\label{eq:lap}
    \mathcal{Z}(\beta)\propto \frac{1}{\Delta(-\beta H)}\lim_{t\to 1}\mathcal{L}_s^{-1}[\tilde{\mathcal{Z}}(\beta,s)](t)
\end{align}
where
\begin{align}\label{eq:dummy}
    \tilde{\mathcal{Z}}(\beta,s)= \frac{1}{d!}
    \int_0^\infty
     D\rho \ \text{det} \ \big[e^{-r_\nu(\beta E_\mu+s)}\big]_{0\leq \nu,\mu\leq d-1}\ 
     \text{det}[r_\nu^{\mu} ]_{0\leq \nu,\mu\leq d-1}.
\end{align}
Here $D\rho=dr_0\cdots d r_{d-1}$ and the domain is extended from $[0,1]$ to $[0,\infty)$.

Now use Andr\'eief’s integral formula \cite{forrester2019meet} with
\begin{align}
    f_\mu(x)=e^{-x (\beta E_\mu+s)},
    \qquad
    g_\mu(x)=x^{\mu}.
\end{align}
Then
\begin{align}
    \braket{f_\mu}{g_\nu}=\int_0^\infty dx \ e^{-x (\beta E_\mu+s)}x^{\nu}
    =\Gamma(\nu+1)\,(\beta E_\mu+s)^{-(\nu+1)},
\end{align}
valid for $\Re(\beta E_\mu+s)>0$. Applying Andr\'eief’s formula gives
\begin{align}
    \tilde{\mathcal{Z}}(\beta,s)
    = \text{det}\Big[\Gamma(\nu+1)\,(\beta E_\mu+s)^{-(\nu+1)}\Big]_{0\leq \nu,\mu\leq d-1}.
\end{align}
Factorising the determinant yields
\begin{align}
    \nonumber
    \tilde{\mathcal{Z}}(\beta,s)
    &= \Bigg(\prod_{\nu=0}^{d-1}\Gamma(\nu+1)\Bigg)
    \ \text{det}\Big[(\beta E_\mu+s)^{-(\nu+1)}\Big]_{0\leq \nu,\mu\leq d-1} \\
    &= \Bigg(\prod_{\nu=0}^{d-1}\Gamma(\nu+1)\Bigg)
    \Bigg(\prod_{\mu=0}^{d-1}(\beta E_\mu+s)^{-d}\Bigg)
    \ \text{det}\Big[(\beta E_\mu+s)^{d-1-\nu}\Big]_{0\leq \nu,\mu\leq d-1}.
\end{align}
The remaining determinant can be put in Vandermonde form,
\begin{align}
    \text{det}\Big[(\beta E_\mu+s)^{d-1-\nu}\Big]_{0\leq \nu,\mu\leq d-1}
    \propto
    \prod_{\nu<\mu}\big[\beta E_\nu-\beta E_{\mu}\big].
\end{align}
Hence
\begin{align}
    \tilde{\mathcal{Z}}(\beta,s)
    \propto 
    \Delta(-\beta H)\,
    \Bigg[\prod_{\mu=0}^{d-1}(\beta E_\mu+s)^{d}\Bigg]^{-1}.
\end{align}
Plugging back into~\eqref{eq:lap} gives
\begin{align}
    \mathcal{Z}(\beta)\propto 
    \lim_{t\to 1}\mathcal{L}^{-1}\bigg[\prod_{k=0}^{d-1}\frac{1}{(\beta E_k+s)^d}\bigg](t).
\end{align}
Using the Bromwich integral,
\begin{align}\label{eq:brom}
    \mathcal{Z}(\beta)\propto \frac{1}{2\pi i}\int^{\gamma+i\infty}_{\gamma-i\infty}ds \ e^{s}\prod_{k=0}^{d-1}\frac{1}{(\beta E_k+s)^{ d}} .
\end{align}
Notice that this expression remains well defined even in the presence of degeneracies. Suppose from now on that there are $D\leq d$ distinct energy levels $E_k\neq E_{k'} \ \forall k\neq k'$, and let integer $n_k$ denote the multiplicity of the $k$'th energy level $E_k$ such that $\sum_{k=1}^{D} n_k=d$. Returning to~\eqref{eq:brom}, we see there are now $D$ simple poles each of order $n_k d$, so by Cauchy's residue theorem we have
\begin{align}\label{eq:part_res}
    \nonumber\mathcal{Z}_{HS}(\beta)&\propto\sum_{k} \text{Res}\bigg[\frac{e^z}{\prod_{j}(\beta E_j+z)^{ n_j d}},-\beta E_k\bigg], \\
    &=\sum_{k} \frac{1}{( n_k d-1)!}\frac{d^{n_k d-1}}{dz^{ n_k d-1}}\bigg(\frac{e^z}{\prod_{j\neq k}(\beta E_j+z)^{ n_j d}}\bigg)\bigg|_{z=-\beta E_k}, 
\end{align}
where summations and products are taken over distinct energy levels. Now we will need to use the Leibniz product rule; for a function $e^{z}\prod_{j\neq k}f_j(z)$ the $(n_k d-1)$'th derivative is
\begin{align}
    \nonumber\frac{d^{n_k d-1}}{dz^{n_k d-1}} \bigg(e^z \prod_{j\neq k}f_j(z)\bigg)&=e^z\sum_{m=0}^{n_k d-1}\binom{n_k d-1}{m} \bigg(\frac{\partial^m }{\partial z^m}\prod_{j\neq k}f_j(z)\bigg), \\
    &=e^z\sum_{m=0}^{n_k d-1}\binom{n_k d-1}{m} \bigg(\sum_{|\vec{m}|=m}m! \prod_{j\neq k}\frac{\partial^{m_j} }{\partial z^{m_j}}\frac{f_j(z)}{m_j!}\bigg)
\end{align}
where $\vec{m}=(m_1,m_2,\cdots m_{D-1})$, $|\vec{m}|=\sum_{j=1}^{D-1} m_j$ and the summation $\sum_{|\vec{m}|=m}$ runs over all $m_j$ from $m_j=0$ to $m_j=m$. Setting $f_j(z)=(\beta E_j+z)^{ -n_j d}$, we also have
\begin{align}
    \frac{\partial^{m_j} }{\partial z^{m_j}}(\beta E_j+z)^{ -n_j d}=\frac{(-1)^{m_j}(n_j d+m_j-1)!}{(n_j d-1)!(\beta E_j+z)^{ n_j d+m_j}},
\end{align}
Therefore combing this with the previous formula we get
\begin{align}
    \frac{d^{n_k d-1}}{dz^{n_k d-1}} \bigg(e^z \prod_{j\neq k}f_j(z)\bigg)\bigg|_{z=-\beta E_k}=e^{-\beta E_k}\sum_{m=0}^{n_k d-1}\binom{n_k d-1}{m} \bigg(\sum_{|\vec{m}|=m}m! \prod_{j\neq k}\binom{n_j d+m_j-1}{m_j}\frac{(-1)^{m_j}}{(\beta E_j-\beta E_k)^{n_j d+m_j}}\bigg)
\end{align}
The final step is to substitute this into~\eqref{eq:part_res}, and after simplifying the factorials we end with
\begin{align}
    \nonumber\mathcal{Z}_{HS}(\beta)&\propto  \sum_{k}\sum_{m=0}^{n_k d-1} \frac{ e^{-\beta E_k}}{(n_k d-m-1)!} \bigg(\sum_{|\vec{m}|=m} \prod_{j\neq k}\binom{n_j d +m_j-1}{m_j}\frac{(-1)^{m_j}}{(\beta E_j-\beta E_k)^{n_j d+m_j}}\bigg), \\
    &=\sum_{k}\sum_{m=0}^{n_k d-1} \frac{(-1)^m e^{-\beta E_k}}{\beta^{m+d(d-n_k)}(n_k d-m-1)!} \bigg(\sum_{|\vec{m}|=m} \prod_{j\neq k}\binom{n_j d +m_j-1}{m_j}\frac{1}{( E_j- E_k)^{n_j d+m_j}}\bigg)
\end{align}
where we used $\sum_{j\neq k}n_j=d-n_k$. This completes the derivation of~\eqref{eq:HSpart2} from main text. 

\section{Density of states: Bures-Hall measure}\label{app:D}

\

Replacing the Hilbert-Schmidt measure with the Bures-Hall measure, the function~\eqref{eq:dummy} becomes
\begin{align}\label{eq:lap2}
    \tilde{\mathcal{Z}}(\beta,s)=\frac{(-1)^{(d-d^2)/2}}{\Delta(\beta H)}\int_{0\leq r_0\leq...\leq  r_{d-1}\leq \infty} 
    D\rho \ \text{det} \ \big[m_\mu( r_\nu,\beta,s)\big]_{\nu,\mu=0,...,d-1}\prod_{\nu<\mu} \frac{ r_\mu- r_\nu}{ r_\nu+ r_\mu},
\end{align}
with
\begin{align}
m_\mu( r_\nu,\beta,s):=\frac{e^{- r_\nu(  \beta E_\mu+s)}}{\sqrt{ r_\nu}}; \ \ \ \ \ 0\leq \nu,\mu\leq d-1.
\end{align}
As before, the integration extends to $[0,\infty)$ after introducing $\delta[\text{Tr}(D)-t]$ prior to Laplace transforming. Using Schur's Pfaffian identity \cite{okada2019pfaffian},
\begin{align}
    \prod_{0\leq\nu<\mu\leq d-1} \frac{ r_\mu- r_\nu}{ r_\nu+ r_\mu}=\begin{cases}
    \text{Pf}[( r_\mu- r_\nu)/( r_\nu+ r_\mu)]_{\nu,\mu=0,...,d-1}, & d \ \text{even}, \\
    \text{Pf}\left[\begin{array}{cc}
          [( r_\mu- r_\nu)/( r_\nu+ r_\mu)]_{\nu,\mu=0,...,d-1} & [1]_{\nu=0,...,d-1}   \\ 
          [-1]_{\mu=0,...,d-1} & 0 
    \end{array}\right], & d \ \text{odd}, \\
    \end{cases}
\end{align}
and de Brujin’s integration theorem \cite{de1955some}, one arrives at
\begin{align}\label{eq:partition4}
    \tilde{\mathcal{Z}}(\beta,s)=(-1)^{d/2}\frac{(-1)^{(d-d^2)/2}}{\Delta(\beta H)}  \text{Pf}\big[A_{\nu,\mu}(\beta,s)\big]_{\nu,\mu=0,...,d-1},
\end{align}
for even $d$, with the standard odd-$d$ augmentation understood. Here
\begin{align}\label{eq:debrujin}
    A_{\nu,\mu}(\beta,s):=\int^\infty_0 dx\int^\infty_0 dy \ m_\nu(x,\beta,s)m_\mu(y,\beta,s)\frac{x-y}{x+y}
\end{align}
for $0\leq \nu,\mu\leq d-1$. For odd $d$ one adjoins an extra row and column indexed by $d$, with
\begin{align}
    A_{\nu,d}(\beta,s)=-A_{d,\nu}(\beta,s):=\int_0^\infty dx\, m_\nu(x,\beta,s),
    \qquad A_{d,d}(\beta,s):=0.
\end{align}
These integrals evaluate to
\begin{align}
    &A_{\nu,\mu}(\beta,s)=\frac{\pi}{\sqrt{(\beta E_\mu+s)(\beta E_\nu+s)}}\bigg(\frac{\sqrt{\beta E_\mu+s}-\sqrt{\beta E_\nu+s}}{\sqrt{\beta E_\mu+s}+\sqrt{\beta E_\nu+s}}\bigg), \\
    &A_{\nu,d}(\beta,s)=-\frac{\sqrt{\pi}}{\sqrt{\beta E_\nu+s}}.
\end{align}
Using Schur's Pfaffian identity again then gives
\begin{align}\label{eq:partition5}
    \tilde{\mathcal{Z}}(\beta,s)=\pi^{d/2}\bigg(\prod^{d-1}_{k=0}\frac{ 1 }{\sqrt{\beta E_k+s}}\bigg)\prod_{0\leq \nu<\mu\leq d-1} \bigg(\frac{1}{\sqrt{\beta E_\nu+s}+\sqrt{\beta E_\mu+s}}\bigg)^2 .
\end{align}
Therefore
\begin{align}
    \mathcal{Z}(\beta)=\text{const}. \ \beta^{1-\frac{d^2}{2}}\mathcal{L}^{-1}\bigg[\prod^{d-1}_{\nu,\mu=0}\frac{1}{\sqrt{  E_\nu+s}+\sqrt{  E_\mu+s}}\bigg](\beta).
\end{align}
Introducing shifted energies $\Delta E_k=E_k-E_0$ and
\begin{align}
    G(s)=\text{const}.\prod^{d-1}_{\nu,\mu=0}\frac{1}{\sqrt{\Delta E_\nu+s}+\sqrt{\Delta E_\mu+s}},
\end{align}
we may equivalently write 
\begin{align}
    \mathcal{Z}(\beta)=\text{const}. \ e^{-\beta E_0}\,\beta^{1-\frac{d^2}{2}}\mathcal{L}_s^{-1}[G](\beta).
\end{align}
Inverting this relation yields the generalised Stieltjes transform
\begin{align}
G(s)=\Gamma(1+d^2/2)\int^\infty_0 d E \ \frac{\Omega(E+E_0)}{(s+E)^{\frac{d^2}{2}+1}}.
\end{align}

\section{Inverting the Stieltjes transform}\label{app:E}

\

Begin with the inversion formula and divide the integral on $[0,\Delta E]$ into sub-intervals,
\begin{align}
    \Omega(E)=\sum_{\Delta E_k < \Delta E }\int^{\widetilde{\Delta E}_{k+1}}_{\Delta E_k} ds \ (\Delta E-s)^{\frac{d^2}{2}-1}D_k(s),
\end{align}
where we define $\Delta E_d:=+\infty$, $\widetilde{\Delta E}_k=\text{min}\{\Delta E,\Delta E_k\}$, and
\begin{align}
     D_k(s)&=\frac{1}{2\pi i}\lim_{\delta\to 0^+}\bigg(G(-s-i\delta)-G(-s+i\delta)\bigg). 
\end{align}
Expanding the generating function gives
\begin{align}
    G(-s-i0)=\prod^{d-1}_{j=0}\frac{1}{\sqrt{\Delta E_j-s-i0}}\prod_{\nu<\mu}\bigg(\frac{1}{\sqrt{\Delta E_\nu-s-i0}+\sqrt{\Delta E_\mu-s-i0}}\bigg)^2.
\end{align}
For the first product,
\begin{align}
    \sqrt{\Delta E_j-s\pm i0}=\begin{cases}
        \mp i\sqrt{s-\Delta E_j}, & j\leq k, \\
        \sqrt{\Delta E_j-s}, & j>k.
    \end{cases}
\end{align}
Considering the second product over $\nu<\mu$, there are three cases:
\begin{itemize}
    \item For $\nu,\mu> k$, the denominator is real with no phase contribution.
    \item For $\nu,\mu\leq k$, one picks up the phase corresponding to
    \[
    i(\sqrt{|\Delta E_\nu-s|}+\sqrt{|\Delta E_\mu-s|}),
    \]
    and there are $\binom{k+1}{2}$ such pairs.
    \item For $\nu\leq k<\mu$,
    \[
    \sqrt{\Delta E_\nu-s-i0}+\sqrt{\Delta E_\mu-s-i0}
    =
    \sqrt{|\Delta E_\mu-s|}-i\sqrt{|\Delta E_\nu-s|},
    \]
    with phase contribution
    \[
    -\, \text{arctan}\bigg(\sqrt{\frac{|\Delta E_\nu-s|}{|\Delta E_\mu-s|}}\bigg).
    \]
\end{itemize}
Collecting phases yields
\begin{align}
    \text{Arg} \big(G(-s-i0)\big)=-\frac{(k+1)\pi}{2} - \frac{(k+1)k\pi}{2} -  2\sum_{\nu\leq k<\mu}\text{arctan}\bigg(\sqrt{\frac{s-\Delta E_\nu}{\Delta E_\mu-s}}\bigg)=-\theta_k(s),
\end{align}
where
\begin{align}
    \theta_k(s)=\frac{(k+1)^2 \pi}{2}+2\sum_{\nu\leq k<\mu}\text{arctan}\bigg(\sqrt{\frac{s-\Delta E_\nu}{\Delta E_\mu-s}}\bigg).
\end{align}
Hence
\begin{align}
    \Omega(E)\propto\sum_{ \Delta E_k\leq  \Delta E}\int^{ \widetilde{\Delta E}_{k+1}}_{ \Delta E_k} ds  ( \Delta E-s)^{\frac{d^2}{2}-1}e^{-R_k(s)}\text{sin}  \theta_k(s),
\end{align}
with magnitude
\begin{align}
    \nonumber R_k(s)&=\frac{1}{2}\sum_{j=0}^{d-1} \ln\!\big(\lvert \Delta E_j - s\rvert
      \big)+\sum_{\nu\leq k<\mu} \text{ln}(\Delta E_\mu-\Delta E_\nu) \\
    & \ \ \ \ +\sum_{\substack{\nu<\mu\\(\nu,\mu\le k)\lor(\nu,\mu>k)}}\!
2\ln\!\Bigl(\sqrt{\lvert \Delta E_\nu - s\rvert}
      +\sqrt{\lvert \Delta E_\mu - s\rvert}\Bigr). 
\end{align}

\textit{Two-level system:} As a sanity check one may recover the same formula~\eqref{eq:qubitdos} for the density of states derived using the Bloch parameterisation. For $d=2$ one has $\Delta E_0=0$ and $\Delta E_1=\epsilon$, so
\begin{align}
    \nonumber\Omega(E)&\propto\int^{ E}_{ 0} ds  \frac{( E-s)}{\epsilon\sqrt{s(\epsilon-s)}}\cos\bigg(2 \ \text{arctan}\bigg(\sqrt{\frac{s}{ \epsilon-s}}\bigg)\bigg), \\
    \nonumber&=\int^{ E}_{ 0} ds  \frac{( E-s)(\epsilon-2s)}{\epsilon^2\sqrt{s(\epsilon-s)}}, \\
    &\propto\bigg(\frac{2 E-\epsilon}{\epsilon^2}\bigg)\sqrt{\epsilon E-E^2}+\text{arcsin}\bigg(\sqrt{\frac{E}{\epsilon}}\bigg),
\end{align}
which is the same as~\eqref{eq:qubitdos}.

\textit{Regularisation:} For numerical purposes it is useful to transform the integral by introducing
\begin{align}
    s=\Delta E_k+\Delta_k t^2, \ \ \ \Delta_k=\widetilde{\Delta E}_{k+1}-\Delta E_k, \ \ \  t\in[0,1]
\end{align}
so that
\begin{align}
    \Omega(E)=2\sum_{ \Delta E_k\leq  \Delta E}\int^{ 1}_{0} dt \ \Delta_k t  ( \Delta E-\Delta E_k-\Delta_k t^2)^{\frac{d^2}{2}-1}e^{-R_k(\Delta E_k+\Delta_k t^2)}\text{sin}  \theta_k(\Delta E_k+\Delta_k t^2).
\end{align}
This form is convenient for numerical integration.

\section{Low energy expansion of density of states for non-interacting spin model}\label{app:F}

\

To analyse the behaviour of the density of states close to the ground state energy shell, we first look at the corresponding low temperature (ie. large $\beta\gg 1$) expansion of the canonical partition function. For this we take an asymptotic expansion of the Bromwich integral 
\begin{align}
\nonumber \beta^{\frac{d^2}{2}-1}e^{\beta E_0}\mathcal{Z}(\beta)&\propto \frac{1}{2\pi i}\int^{\gamma+i\infty}_{\gamma-i\infty}ds \ \frac{e^{s\beta}}{\sqrt{s}} \prod^{d-1}_{k=1}\frac{1}{\sqrt{  \Delta E_k+s}}\prod_{\nu<\mu}\bigg(\frac{1}{\sqrt{  \Delta E_\nu+s}+\sqrt{  \Delta E_\mu+s}}\bigg)^2 \\
&\sim \bigg[\prod^{d-1}_{k=1}\frac{1}{\sqrt{  \Delta E_k}}\prod_{\nu<\mu}\bigg(\frac{1}{\sqrt{  \Delta E_\nu}+\sqrt{  \Delta E_\mu}}\bigg)^2\bigg]\bigg[\beta^{-1/2}+\frac{1}{4 \beta^{3/2}}\bigg(\sum_{k=1}^{d-1}\frac{1}{\sqrt{\Delta E_k}}\bigg)^2+\mathcal{O}(\beta^{-5/2})\bigg].
\end{align}
To connect this to the density of states it is convenient to first define a shifted Hamiltonian $H_0=H-E_0 \mathbb{I}\geq 0$ and a corresponding density 
\begin{align}
    \omega_0(E)=\int_{\mathcal{S}(\mathcal{H})}dV(\rho) \ \delta[E-\tr{H_0\rho}]=\omega(E+E_0).
\end{align}
This is related to the original partition function via a Laplace transform,
\begin{align}\label{eq:omega0}
    \mathcal{Z}(\beta)=e^{-\beta E_0} \int^\infty_0 dE \ e^{-\beta E}\omega_0(E)=e^{-\beta E_0} \mathcal{L}_E[\omega_0](\beta).
\end{align}
Next, we propose an ansatz for the asymptotic expansion of $\omega_0(E)$ around $E\sim 0$, 
\begin{align}\label{eq:watson}
    \omega_0(E)\sim E^{\lambda}\sum_{n=0}^{\infty} c_n E^{n}, \ \ \lambda>-1.
\end{align}
By Watson's lemma, the Laplace transform will be
\begin{align}
   \mathcal{L}_E[\omega_0](\beta)\sim \sum_{n=0}^\infty c_n \Gamma(\lambda+n+1) \beta^{-(\lambda+n+1)}.  
\end{align}
Combining this with~\eqref{eq:omega0} we find $\lambda=\frac{d^2-3}{2}$ and therefore the leading-order contributions to $\Omega(E)$ go as
\begin{align}
    \Omega(E)=\int^{E-E_0}_{0} dE' \ \omega_0(E')
    \sim A_0 (E-E_0)^{\frac{d^2-1}{2}}+A_1 (E-E_0)^{\frac{d^2+1}{2}}+\mathcal{O}((E-E_0)^{\frac{d^2+3}{2}}),
\end{align}
with coefficients proportional to
\begin{align}
    A_0\propto \prod_{k=1}^{d-1}\frac{1}{\sqrt{  \Delta E_k}}\prod_{\nu<\mu}\bigg(\frac{1}{\sqrt{  \Delta E_\nu}+\sqrt{  \Delta E_\mu}}\bigg)^2,
\end{align}
\begin{align}
    A_1\propto \bigg[\prod_{k=1}^{d-1}\frac{1}{\sqrt{  \Delta E_k}}\prod_{\nu<\mu}\bigg(\frac{1}{\sqrt{  \Delta E_\nu}+\sqrt{  \Delta E_\mu}}\bigg)^2 \bigg] \bigg(\sum_{k=1}^{d-1}\frac{1}{\sqrt{\Delta E_k}}\bigg)^2.
\end{align}
For the non-interacting spin model this may then be reorganised over the distinct excited levels and multiplicities, leading to the entropy expansion quoted in the main text. To get the final result for the entropy we expand the logarithm,
\begin{align}
    \text{ln}(A x^k+Bx^{k+1}+\mathcal{O}(x^{k+2}))=\text{ln}(A) +k \ \text{ln}(x)+\frac{B}{A}x+\mathcal{O}(x^{2}).
\end{align}

\section{Non-interacting spin model}\label{app:G}

\

We begin with the low-energy approximation of the volume entropy stated in the main text~\eqref{eq:Sexpand}, and apply the perturbation $H\mapsto H+\lambda H^2$. For the non-interacting spin model the energy levels are
\begin{align}
    E_k=-\frac{NB}{2}+kB, \qquad k=0,1,\dots,N,
\end{align}
with multiplicities $\binom{N}{k}$. Under the perturbation, the ground-state energy becomes
\begin{align}
    E_0(\lambda)=E_0+\lambda E_0^2=-\frac{NB}{2}+\lambda\frac{N^2B^2}{4},
\end{align}
and the excited gaps relative to the perturbed ground state are
\begin{align}
    \Delta E_k(\lambda)
    &=E_k(\lambda)-E_0(\lambda) \\
    &=\big(E_k+\lambda E_k^2\big)-\big(E_0+\lambda E_0^2\big) \\
    &=kB+\lambda B^2k(k-N).
\end{align}
Therefore the entropy takes the form
\begin{align}
    \nonumber S_\lambda(E)&\sim \frac{2^{2N}-1}{2}\ln \bigg(E+\frac{NB}{2}-\lambda \frac{N^2B^2}{4}\bigg) \\
    &\quad -\sum_{\nu,\mu=1}^N\binom{N}{\nu} \binom{N}{\mu}\ln\!\bigg(\sqrt{B\nu+\lambda B^2\nu(\nu-N)}+\sqrt{B\mu+\lambda B^2\mu(\mu-N)}\bigg)  \\
    &\quad +\frac{\bigg(E+\frac{N B}{2}-\lambda \frac{N^2 B^2}{4}\bigg)}{(2^{2N+1}-2)}\bigg(\sum_{k=1}^N \binom{N}{k}\frac{1}{\sqrt{k B+\lambda B^2k(k-N)}}\bigg)^2.
\end{align}
To first order in $\Delta E:=E+\frac{NB}{2}$ the temperature of the system is
\begin{align}
T(E)=\frac{(2E+BN)}{2^{2N}-1}+\mathcal{O}(\Delta E^2).
\end{align}

Using the thermodynamic formula~\eqref{eq:pk} we compute the second moment of energy:
\begin{align}
    \frac{\langle H^2 \rangle}{N^2B^2}
    =-\frac{1}{B^2N^2}T(E)\frac{\partial S_\lambda}{\partial\lambda}\bigg|_{\lambda=0}.
\end{align}
Evaluating the derivative term by term gives
\begin{align}
    \nonumber \frac{\langle H^2 \rangle}{N^2B^2}
    &=\frac{1}{4}
    +\frac{\big(\bar{\epsilon}/B+1/2\big)}{N(2^{2N}-1)}
    \bigg[
        N(2^N-1)
        -\bigg(\sum_{k=1}^N\binom{N}{k}\sqrt{k}\bigg)^2
    \bigg] \\
    &\quad
    +\frac{\big(\bar{\epsilon}/B+1/2\big)}{4(2^{2N}-1)^2}
    \bigg(\sum_{k=1}^N\binom{N}{k}\frac{1}{\sqrt{k}}\bigg)^2,
\end{align}
where $\bar{\epsilon}=E/N$.

The relative fluctuations then become
\begin{align}
    \nonumber \Delta \bar{\epsilon}_N^2
    &=\frac{\langle H^2\rangle-E^2}{E^2} \\
    &=\frac{B^2}{4\bar{\epsilon}^2}-1
    +\frac{B(\bar{\epsilon}+B/2)}{\bar{\epsilon}^2}
    \bigg[
        \frac{N(2^N-1)-\big(\sum_{k=1}^N\binom{N}{k}\sqrt{k}\big)^2}{N(2^{2N}-1)}
        +\frac{1}{4(2^{2N}-1)^2}
        \bigg(\sum_{k=1}^N\binom{N}{k}\frac{1}{\sqrt{k}}\bigg)^2
    \bigg].
\end{align}

To analyse the two remaining sums for large $N$, introduce the binomially distributed random variable $K\sim \mathrm{Binom}(N,1/2)$ and denote the corresponding average by $\mathbb{E}[\,\cdot\,]$. One may then express the summations as
\begin{align}
    \sum_{k=1}^N\binom{N}{k}\sqrt{k}
    &=2^N\mathbb{E}[\sqrt{K}]
    =2^N \sqrt{\frac{N}{2}}\bigg(1-\frac{1}{8N}+\mathcal{O}(1/N^2)\bigg), \\
    \sum_{k=1}^N\binom{N}{k}\frac{1}{\sqrt{k}}
    &=2^N\mathbb{E}[K^{-1/2}]
    =2^N \sqrt{\frac{2}{N}}\bigg(1+\frac{3}{8N}+\mathcal{O}(1/N^2)\bigg).
\end{align}
Plugging this into $\Delta \bar{\epsilon}_N^2$ and retaining the leading terms gives
\begin{align}
    \Delta \bar{\epsilon}_N^2
    =-1-\frac{B}{2\bar{\epsilon}}
    +\frac{B(B+2\bar{\epsilon})}{16N\,\bar{\epsilon}^2}
    +\mathcal{O}(1/N^2).
\end{align}
We therefore conclude that the relative fluctuations approach a finite value in the thermodynamic limit,
\begin{align}
    \Delta \bar{\epsilon}_\infty^2=-1-\frac{B}{2\bar{\epsilon}},
\end{align}
which is non-negative on the physical interval $-B/2\leq \bar{\epsilon}<0$ and vanishes only when the system reaches the ground state. This gives the  fluctuation formulas stated in the main text.

\end{document}